%% file: main.tex
\documentclass[letterpaper,hidelinks,journal,10pt]{IEEEtran}
\IEEEoverridecommandlockouts

\usepackage[utf8]{inputenc}
\usepackage[T1]{fontenc}

\usepackage[noadjust]{cite}
\usepackage{booktabs}

\usepackage[caption=false]{subfig}
\usepackage{amsmath}
\usepackage{amssymb}
\usepackage{physics}

\usepackage{float}
\usepackage{capt-of}

\usepackage{tabularx}
\newcolumntype{Y}{>{\centering\arraybackslash\hsize=1.1\hsize}X}
\newcolumntype{Z}{>{\hsize=0.6\hsize}X}
\usepackage{dblfloatfix}

\usepackage{url}
\usepackage[colorlinks=false]{hyperref}
\usepackage[nameinlink,capitalise]{cleveref}
\usepackage{bm}
\usepackage{tikz}
\usepackage{circuitikz}
\usetikzlibrary{shapes,arrows,calc,fit,matrix,quotes,positioning}
\usepackage[ruled,vlined]{algorithm2e}
\SetKwInput{kwInit}{Initialise}

\makeatletter
\newcommand{\removelatexerror}{\let\@latex@error\@gobble}
\makeatother

\usepackage{dsfont}

\usepackage{epstopdf}
\makeatletter
\def\subsubsection{%
  \@startsection
    {subsubsection}                 % type
    {3}                             % level
    {\parindent}                    % indent
    {1.5ex plus 1.5ex minus 1.5ex}  % beforeskip {0ex plus 0.1ex minus 0.1ex}
    {0.7ex plus .5ex minus 0ex}     % afterskip {0ex}
    {\normalfont\normalsize\itshape}% style
}
\makeatother

\usepackage{etoolbox}
\makeatletter
\patchcmd{\@algocf@start}% <cmd>
  {-1.5em}% <search>
  {0pt}% <replace>
  {}{}% <success><failure>
\makeatother

\title{Learning-Based Reconstruction of FRI Signals}

\author{Vincent C. H. Leung, \IEEEmembership{Member,~IEEE,} Jun-Jie Huang, \IEEEmembership{Member,~IEEE,} \\Yonina C. Eldar, \IEEEmembership{Fellow,~IEEE,} and Pier Luigi Dragotti, \IEEEmembership{Fellow,~IEEE}
\thanks{Vincent C. H. Leung and Pier Luigi Dragotti are with the Department of Electrical and Electronic Engineering, Imperial College London, United Kingdom. Email: \{\href{mailto:chi.leung14@imperial.ac.uk}{chi.leung14}, \href{mailto:p.dragotti@imperial.ac.uk}{p.dragotti}\}@imperial.ac.uk}
\thanks{Jun-Jie Huang is with the College of Computer Science, National University of Defense Technology, China. Email: \href{mailto:jjhuang@nudt.edu.cn}{jjhuang@nudt.edu.cn}}
\thanks{Yonina C. Eldar is with the Faculty of Mathematics and Computer Science, Weizmann Institute of Science, Rehovot, Israel. Email: \href{mailto:yonina.eldar@weizmann.ac.il}{yonina.eldar@weizmann.ac.il}}
}

\begin{document}
%\ninept
\bstctlcite{IEEEexample:BSTcontrol}
\maketitle
\begin{abstract}
Finite Rate of Innovation (FRI) sampling theory enables reconstruction of classes of continuous non-bandlimited signals that have a small number of free parameters from their low-rate discrete samples. This task is often translated into a spectral estimation problem that is solved using methods involving estimating signal subspaces, which tend to break down at a certain peak signal-to-noise ratio (PSNR). To avoid this breakdown, we consider alternative approaches that make use of information from labelled data. We propose two model-based learning methods, including deep unfolding the denoising process in spectral estimation, and constructing an encoder-decoder deep neural network that models the acquisition process. Simulation results of both learning algorithms indicate significant improvements of the breakdown PSNR over classical subspace-based methods. While the deep unfolded network achieves similar performance as the classical FRI techniques and outperforms the encoder-decoder network in the low noise regimes, the latter allows to reconstruct the FRI signal even when the sampling kernel is unknown. We also achieve competitive results in detecting pulses from in vivo calcium imaging data in terms of true positive and false positive rate while providing more precise estimations.
\end{abstract}
\begin{IEEEkeywords}
Finite rate of innovation, model-based neural networks, autoencoders, deep unfolding, signal reconstruction, deep learning.
\end{IEEEkeywords}
\IEEEpeerreviewmaketitle

\section{Introduction}
\label{sect:intro}
\IEEEPARstart{C}{lassical} sampling theory enables perfect reconstruction of continuous shift-invariant signals from their discrete samples \cite{Eldar2014}. In recent years, the emergence of finite rate of innovation (FRI) sampling theory \cite{Vetterli2002,Dragotti2007,Blu2008,Uriguen2013,Tur2011} has extended sampling results to classes of non-bandlimited signals that have finite degrees of freedom per unit time. The most basic FRI signal is a stream of $K$ pulses, which has a $2K$ rate of innovation as the signal can be defined by the amplitudes and the locations of $K$ pulses. This has led to a wide range of applications such as calcium imaging \cite{Onativia2013}, functional magnetic resonance imaging (fMRI) \cite{Dogan2014}, radar \cite{Bar-Ilan2014}, ultrasound imaging \cite{Wagner2012} and electrocardiogram (ECG) \cite{Hao2005}.

The existing FRI signal reconstruction algorithms usually transform the continuous location estimation problem into an exponential frequency estimation problem which can be solved by spectral estimation techniques such as Prony's method with Cadzow denoising \cite{Prony1795,Cadzow1988} and matrix pencil \cite{Hua1990}. These methods involve signal subspace estimation by performing the Singular Value Decomposition (SVD) to estimate signal subspaces. Under noisy conditions, the reconstruction performance follows the Cram\'{e}r-Rao bound in the low noise regime \cite{Cramer1946,Rao1945}. However, it breaks down when the peak signal-to-noise ratio (PSNR) drops below a certain threshold. The reason is conjectured to be the so-called subspace swap event \cite{Wei2015} which refers to the confusion of the orthogonal subspace with the signal subspace under noisy conditions \cite{Thomas1995}.

In addition to classical FRI techniques, compressed sensing (CS) also allows to reconstruct the locations of a stream of pulses on a grid \cite{Bar-Ilan2014}. However, both CS and FRI approaches involve finding the Fourier coefficients of the sampling kernel at certain frequencies \cite{Uriguen2013}, which means that the sampling kernel has to be known in order to reconstruct the signal. In many practical applications such as calcium imaging in neuroscience \cite{Onativia2013}, the information of the sampling kernel is unknown. While extensions to CS have enabled off-the-grid reconstruction of continuous-time streams of pulses using atomic norm \cite{Bhaskar2013,Tang2013}, or convex relaxation \cite{Candes2014}, they still suffer from similar drawbacks as standard FRI: the pulse shape must be known a priori. Alternatively, compressive multichannel blind deconvolution \cite{Mulleti2020, Tolooshams2022} allows to reconstruct FRI signals from the low-rate samples acquired by multiple kernels without knowing their shapes. Nonetheless, in this paper, we consider the case of having only a single pulse shape.

Here we aim to overcome these limitations by adopting data-driven learning-based approaches. Several existing works utilise deep neural networks to perform spectral estimation on problems such as estimating the frequencies of multisinusoidal signals \cite{Mathew1994,Izacard2019,Izacard2019a}, or estimating the direction of arrival of multiple sound sources \cite{Adavanne2018,Xiao2015}. In this paper, we focus on developing interpretable networks \cite{Shlezinger2021} that are based on the existing FRI reconstruction model.

To address the performance breakdown, we begin by proposing deep unfolding the denoising process that is used before solving the transformed frequency estimation problem in classical FRI methods. Deep unfolding \cite{Gregor2010,Monga2021,Solomon2020} maps iterative algorithms into layers of networks with learnable parameters while keeping the domain knowledge of the data, in our case, the spectral sample matrix being Toeplitz and low rank. Here, we choose to unfold the projected Wirtinger gradient descent (PWGD) algorithm \cite{Cai2015a}, which is a slight variation of Cadzow denoising that allows to embed the learnable parameters into the network. The reconstructed locations of the Diracs are then obtained after coupling the unfolded network with Prony's method. We use the zero eigenvalue-based loss function proposed in \cite{Dang2018}, which aims to minimise the projection denoised matrix along the directions of the ground truth annihilating filter, while maximising the projection along the orthogonal subspace. This reduces the occurrence of subspace swap events and thus improves the breakdown PSNR.

Alternatively, as transforming from the continuous location estimation problem to exponential frequency problem still requires knowledge of the sampling kernel, we propose bypassing spectral estimation by learning an encoder network $g_\phi(\cdot)$ that aims to infer the locations of the Diracs directly from the noisy samples. We then fine-tune by appending a decoder network $f_\theta(\cdot)$ that models the acquisition process for FRI signals to resynthesise the samples based on the estimated locations and amplitudes. Together, they form FRI Encoder-Decoder Network (FRIED-Net). Depending on whether the sampling kernel is known, we can either fix the parameters of the decoder $\theta$ or make them learnable. This addresses the application to calcium imaging when the sampling kernel is unknown and hence classical methods fail to reconstruct the FRI signal. The loss function considers the error on both reconstructed locations and the corresponding resynthesised discrete samples, as the latter provides a regularising effect on the output of the encoder network. 

Using a stream of $K$ Diracs as an example, we demonstrate how the two proposed model-based deep learning approaches can reconstruct the locations of the Diracs. The contributions of this paper is as follows.
\begin{itemize}
    \item Both deep unfolded Wirtinger gradient descent and FRIED-Net overcome the breakdown event, regardless of whether the sampling kernel is known or not. We show this by comparing the performance of classical FRI methods and our proposed learning-based techniques for reconstructing a stream of $K$ Diracs.
    \item Our proposed FRIED-Net can reconstruct the FRI signal without knowing the sampling kernel, and is capable of learning it. While classical FRI methods normally require knowledge of the sampling kernel, our approach can reconstruct FRI signals by inferring the FRI parameters directly from the noisy samples using neural networks. Furthermore, we can learn the kernel, which is represented by the parameters of the decoder $\theta$, through backpropagation without affecting significantly the reconstruction performance. We show this by applying FRIED-Net to spike detection in calcium imaging data. 
\end{itemize}
This paper is an extension to our previous work presented in \cite{Leung2021}. 

The rest of the paper is organised as follows: In \Cref{sect:classicalfri}, we discuss the inherent subspace swap event and breakdown PSNR in classical FRI methods using the example of reconstructing a stream of $K$ Diracs. We then present our proposed learning-based FRI reconstruction approaches in \Cref{sect:learningbasedFRI}. In \Cref{sect:simulation}, we compare our approaches against the classical FRI techniques under different settings. In \Cref{sect:application}, we apply FRIED-Net to detect spikes from calcium imaging data. We then conclude in \Cref{sect:conclusion}.

\section{Classical FRI Reconstruction Methods}
\label{sect:classicalfri}

\begin{figure}[t]
\centering
\input{figures/acquisition.tex}
\caption{Acquisition process that converts continuous time signal $x(t)$ into discrete time samples $y[n]= \left\langle x(t),\varphi\left({t}/{T}-n\right)\right\rangle$.}
\label{fig:acquisition}
\end{figure}
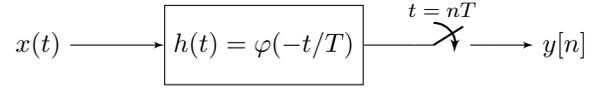

In this section, we overview classical FRI reconstruction methods and explain the breakdown event in noisy conditions using an example of a stream of Diracs. \cref{fig:acquisition} illustrates a typical acquisition process that involves filtering the input continuous-time signal $x(t)$ with $h(t)=\varphi(-t/T)$ and sampling at a regular interval $t=nT$. Perfect reconstruction of classes of FRI signals can be achieved by using specific classes of sampling kernels $\varphi(t)$ (e.g. \cite{Vetterli2002,Dragotti2007,Uriguen2013,Tur2011}). 

For example, we can consider the reconstruction of a $\tau$-periodic stream of $K$ Diracs: 
\begin{align}
  x(t) = \sum_{l\in\mathbb{Z}}\sum_{k=0}^{K-1} a_k \delta(t-t_k-l\tau)\text{,}
\end{align}
where $\{a_k\in \mathbb{R}\}_{k=0}^{K-1},\{t_k\in \mathbb{R}\}_{k=0}^{K-1}$ are the amplitudes and locations of the Diracs respectively. One of the sampling kernels $\varphi(t)$ that allows to reconstruct a stream of Diracs is the exponential reproducing function which, together with its uniform shifts weighted by proper coefficients $c_{m,n}$, can reproduce complex exponentials \cite{Unser2005}:
\begin{align}
    \sum_{n\in\mathbb{Z}}c_{m,n}\varphi(t-n) = e^{j\omega_m t}\text{,} \label{eq:exponentialreproducing}
\end{align}
with frequencies $\omega_m = \omega_0 +m\lambda$ for $m = 0,1,...,P$, where $\lambda \in \mathbb{R}$ is the separation between the equispaced $\omega_m$. Using \eqref{eq:exponentialreproducing} and assuming a sampling period $T = \tau/N$, it is possible to map the acquired samples  
\begin{align}
    y[n] = \left\langle x(t),\varphi\left(\frac{t}{T}-n\right)\right\rangle = \sum_{k=0}^{K-1} a_k \varphi\left(\frac{t_k}{T}-n\right) \text{,} \label{eq:samples}
\end{align}
into a sum of exponentials:
\begin{align}
    s[m] & = \sum_{n=0}^{N-1}c_{m,n}y[n] = \sum_{k=0}^{K-1}a_k \sum_{n\in\mathbb{Z}}c_{m,n}\varphi\left(\frac{t_k}{T}-n\right) \nonumber \\
    & = \sum_{k=0}^{K-1}\underbrace{a_k e^{j\omega_0t_k/T}}_{b_k} \left(\underbrace{e^{j\lambda t_k/T}}_{u_k}\right)^m = \sum_{k=0}^{K-1} b_k u_k^m\text{.} \label{eq:nonlinear}
\end{align}
The amplitudes of the Diracs $\left\{a_k\right\}_{k=0}^{K-1}$ are mapped to the amplitudes of the exponentials $\{b_k\}_{k=0}^{K-1}$ while the locations of Diracs $\{t_k \}_{k=0}^{K-1}$ are transformed to $\{u_k \}_{k=0}^{K-1}$. This forms a spectral estimation problem. In this paper, we are particularly interested in retrieving the locations of the Diracs $\{t_k\}_{k=0}^{K-1}$ due to its non-linear nature in the problem seen in \eqref{eq:nonlinear}. The problem of retrieving the amplitudes of the Diracs is linear, which means that given the locations, we can directly estimate the amplitudes. We also note that information of the sampling kernel $\varphi(t)$ is implicitly included in the coefficients $c_{m,n}$.

One of the most common techniques to solve the spectral estimation problem is Prony's method \cite{Prony1795}. It shows that there exists a filter $\mathbf{h}$ of length $K+1$ that annihilates the sequence $s[m]$, i.e. $s[m]*h[m]=0$, and the roots of this annihilating filter give us $\{u_k \}_{k=0}^{K-1}$. To find the coefficients of $\mathbf{h}$, we rewrite the convolution into matrix form:
\begin{align}
    \mathbf{Sh}= 
    \begin{bmatrix}
    s[K] & s[K-1] & \dots & s[0]\\
    s[K+1] & s[K] & \dots & s[1]\\
    \vdots & \vdots & \ddots & \vdots\\
    s[P] & s[P-1] & \dots & s[P-K]\\
    \end{bmatrix}
    \begin{bmatrix}
    1\\
    h[1]\\
    \vdots\\
    h[K]
    \end{bmatrix}=\mathbf{0}.
\end{align}
Since $\mathbf{S} \in \mathbb{C}^{(P-K+1)\times(K+1)}$ is of rank-$K$ and $\mathbf{h}$ lies in the nullspace, we can obtain $\mathbf{h}$ by performing a singular value decomposition (SVD) on $\mathbf{S}$ and choosing the right singular vector corresponding to the zero singular value.

\subsection{Reconstruction under Noisy Conditions}
\label{sect:noisyfri}
Often the acquisition process induces noise. The noisy samples can be written as
\begin{align}
    \tilde{y}[n]=y[n]+\varepsilon[n], \label{eq:noisyyn}
\end{align}
where $\varepsilon[n]$ is additive white Gaussian noise with standard deviation $\sigma_\varepsilon$. Since the matrix $\tilde{\mathbf{S}}$ is now noisy and hence full rank, the nullspace is trivial and we instead estimate the annihilating filter by finding the right singular vector with the smallest singular value.

Furthermore, we can make Prony's method more resilient to noise by cleaning the observed sum of exponentials. Since the ideal noiseless matrix $\mathbf{S}$ is of rank $K$ and Toeplitz, we aim to find a denoised matrix $\hat{\mathbf{S}}$ that is closest to the noisy matrix $\tilde{\mathbf{S}}$ while possessing these two properties. This is also known as structured low rank approximation (SLRA) \cite{Markovsky2008}.

SLRA can be solved by using the classical iterative Cadzow denoising algorithm \cite{Cadzow1988}, which performs alternating projections between the set of rank-$K$ matrices and the set of Toeplitz matrices, denoted by $\mathcal{P}_{\mathcal{R}_K}(\cdot)$ and $\mathcal{P}_{\mathcal{T}}(\cdot)$ respectively. The former is done by performing SVD and keeping the $K$ largest singular values (hard thresholding), while the latter is done by averaging each diagonal of the matrix. As the algorithm performs better when the Toeplitz matrix is near square, we start with constructing a similar Toeplitz matrix $\tilde{\mathbf{S}}_M \in \mathbb{C}^{(P-M+1)\times (M+1)}$, where $M=\lceil P/2\rceil$, before reshaping the denoised matrix $\hat{\mathbf{S}}_M$ into $\hat{\mathbf{S}} \in \mathbb{C}^{(P-K+1)\times(K+1)}$ to apply Prony's method.

\subsection{Breakdown PSNR}
\label{sect:breakdownpsnr}

\begin{figure}[t]
    \centering
    \includegraphics[width=0.9\linewidth]{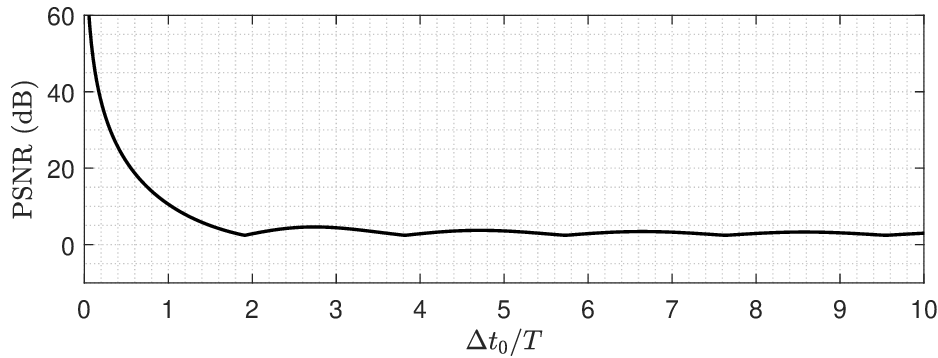}
    \vspace*{-0.2cm}
    \caption{Relationship between breakdown PSNR and the distance between Diracs in the case of $K = 2, N = P+1 = 21 \text{ and } \lambda = \frac{2\pi}{P+1}$ (after \cite{Wei2015}). The subspace-based methods will break down in the region below the curve.}
    \label{fig:breakdownPSNR}
    \vspace*{-0.2cm}
\end{figure}

Despite the fact that Cadzow denoising helps the classical subspace-based methods achieve optimal reconstruction performance defined by the Cram\'{e}r-Rao bound, previous works such as \cite{Blu2008} have shown that they break down at a certain PSNR threshold. It is conjectured that the breakdown in subspace-based techniques is due to the confusion between noise and signal subspaces in performing spectral estimation \cite{Thomas1995}. In \cite{Wei2015}, a mathematical relationship was drawn between the breakdown PSNR and the relative distance $\Delta t_k/T$ between neighbouring Diracs with $\Delta t_k = t_{k+1} - t_{k}$. For instance, when there is a stream of two Diracs of the same amplitudes $(K=2, a_0=a_1)$ sampled by an exponential reproducing kernel $\varphi(t)$ of maximum-order and minimum-support (eMOMS) \cite{Uriguen2013} that can reproduce $P+1=N$ exponentials, a necessary condition for subspace swap event is
\begin{align}
    \text{PSNR} < 10\log_{10}\frac{8\left(\frac{P}{2} + 1\right) \ln\left(\frac{P}{2} +1\right)}{\left(\frac{P}{2}+1-\frac{\sin(\frac{\lambda}{2}(\frac{P}{2}+1)\Delta t_0/T)}{\sin(\frac{\lambda}{2}\Delta t_0/T)}\right)^2}\text{.} \label{eq:breakdownPSNR}
\end{align}
This is visualised in \cref{fig:breakdownPSNR}, which shows that the smaller the distance between two nearby Diracs, the higher the breakdown PSNR will be. Thus, the subspace swap event suggests that current FRI techniques preclude us from recovering FRI signals with high resolution under strong noise. 

\section{Learning-based FRI Reconstruction}
\label{sect:learningbasedFRI}
To address the breakdown, in this section, we introduce two learning-based FRI reconstruction approaches: \emph{Deep Unfolded Projected Wirtinger Gradient Descent} and \emph{FRI Encoder-Decoder Network (FRIED-Net)}. While the former aims to improve the denoising process of the frequency estimation problem to reduce the occurrence of subspace swap events in classical FRI methods, the latter considers the original FRI reconstruction problem and allows to reconstruct without knowledge of the pulse shape.

\subsection{Deep Unfolded Projected Wirtinger Gradient Descent}
\begin{figure}[!b]
\vspace*{-0.5cm}
\removelatexerror% Nullify \@latex@error
\begin{algorithm}[H]
\DontPrintSemicolon
\SetAlgoLined
\KwIn{$\mathbf{L}^{(0)}=\mathbf{0},\mathbf{H}^{(0)}=\tilde{\mathbf{S}}_M$}
\KwOut{Denoised Toeplitz matrix $\hat{\mathbf{S}}_M=\mathbf{H}^{(L)}$}
Choose the parameters $\delta_1, \delta_2 \in (0,1]$.\;
\For{$l\gets0$ \KwTo $L-1$}{
    $\mathbf{L}^{(l+1)}=\mathcal{P}_{\mathcal{R}_K}\left((1-\delta_1)\mathbf{L}^{(l)}+\delta_1\mathbf{H}^{(l)}\right)$\;
    $\mathbf{H}^{(l+1)}=\mathcal{P}_{\mathcal{T}}\left(\delta_2\mathbf{L}^{(l+1)}+(1-\delta_2)\mathbf{H}^{(l)}\right)$\;
}
\caption{Projected Wirtinger Gradient Descent \cite{Cai2015a}}
\label{algo:wirtinger}
\end{algorithm}
\vspace*{-0.5cm}
\end{figure}

\begin{figure*}[!ht]
\vspace*{-0.2cm}
\centering
\removelatexerror% Nullify \@latex@error
\begin{minipage}{0.49\textwidth}
    \begin{algorithm}[H]
    \DontPrintSemicolon
    \SetAlgoLined
    \KwIn{$\mathbf{L}^{(0)}=\mathbf{0},\mathbf{H}^{(0)}=\tilde{\mathbf{S}}_M$}
    \KwOut{Denoised Toeplitz matrix $\hat{\mathbf{S}}_M=\mathbf{H}^{(L)}$}
    \kwInit{$\begin{aligned}[t] 
    \forall l \in [0, L-1]\colon & \delta_1, \delta_2, \mu^{(l)} \in (0,1], \\
    & \mathbf{W}_1^{(l)} = (1-\delta_1)\mathds{I},\mathbf{W}_2^{(l)} = \delta_1\mathds{I}, \\
    &\mathbf{W}_3^{(l)} = \delta_2\mathds{I}, \mathbf{W}_4^{(l)} = (1-\delta_2)\mathds{I} \\
    \end{aligned}$}
    \For{$epoch\gets0$ \KwTo $N_{epoch}-1$}{
     \For{$l\gets0$ \KwTo $L-1$}{
        $\mathbf{L}^{(l+1)}=\mathcal{S}_{\mu^{(l)}{\sigma_{K+1}}}\left(\mathbf{W}_1^{(l)}\mathbf{L}^{(l)}+\mathbf{W}_2^{(l)}\mathbf{H}^{(l)}\right)$\;
        $\mathbf{H}^{(l+1)}=\mathcal{P}_{\mathcal{T}}\left(\mathbf{W}_3^{(l)}\mathbf{L}^{(l+1)}+\mathbf{W}_4^{(l)}\mathbf{H}^{(l)}\right)$\;
     }
     Update $\left\{\mathbf{W}_1^{(l)},\mathbf{W}_2^{(l)},\mathbf{W}_3^{(l)},\mathbf{W}_4^{(l)},\mu^{(l)}\right\}$ for all $L$ layers by backpropagating the loss function $\mathcal{L}(\mathbf{\hat{S}})$\;
    }
     \caption{Deep Unfolded PWGD}
    \end{algorithm}
\end{minipage}
\hfil
\begin{minipage}{0.5\textwidth}
\input{figures/unfoldedwirtinger.tex}
\end{minipage}
\caption{Deep unfolded projected Wirtinger gradient descent algorithm and its network architecture. Since in our setting each layer has its own set of learnable parameters, the parameters of the $l$-th layer are denoted by $\left\{\mathbf{W}_1^{(l)},\mathbf{W}_2^{(l)},\mathbf{W}_3^{(l)},\mathbf{W}_4^{(l)},\mu^{(l)}\right\}$ and highlighted in red in the block diagram.}
\label{fig:unfoldedwirtinger}
\end{figure*}
We first propose to perform deep unfolding on the denoising process prior to Prony's method. Algorithm unfolding is a technique that aims to convert iterative algorithms into interpretable deep neural networks \cite{Monga2021}. By making the parameters used in the algorithm learnable via backpropagation using training data, the unfolded deep network effectively performs as a parameter-optimised algorithm. 

As mentioned in \cref{sect:noisyfri}, the most common iterative denoising algorithm is Cadzow denoising which alternately projects between the set of rank-$K$ matrices and the set of Toeplitz matrices. A generalised version of Cadzow denoising, projected Wirtinger gradient descent (PWGD) \cite{Cai2015a}, is introduced in Algorithm \ref{algo:wirtinger}.

Similar to Cadzow, PWGD alternately projects between the set of rank-$K$ matrices and the set of Toeplitz matrices. However, there exists constants $\delta_1, \delta_2 \in (0,1]$ that weight averages between the matrices, which can be transformed into learnable parameters. Note that Cadzow denoising can be effectively viewed as a special case of PWGD when $\delta_1=\delta_2=1$. To perform the unfolding, we replace each of the constants $\{(1-\delta_1), \delta_1, \delta_2, (1-\delta_2)\}$ with learnable weight matrices $\left\{\mathbf{W}_1,\mathbf{W}_2,\mathbf{W}_3,\mathbf{W}_4\right\}\in \mathbb{C}^{(P-M+1)\times (P-M+1)}$. 

To add further freedom into the network, we replace the rank-$K$ constraint with its convex surrogate \cite{Fazel2001}, that is the nuclear norm $\|\hat{\mathbf{S}}_M\|_*$. Essentially, we are soft thresholding the singular values of $\hat{\mathbf{S}}_M$ instead of hard thresholding. While this choice of the threshold can often be problematic \cite{Condat2015}, with the aid of unfolding, we can make the threshold learnable via backpropagation. We denote this proximal mapping corresponding to the nuclear norm as $\mathcal{S}_{\mu\sigma_{K+1}}(\mathbf{X})$, which refers to soft thresholding the singular values of $\mathbf{X}$ with threshold $\mu\sigma_{K+1}$. This operation can also be expressed in terms of rectified linear unit (ReLU) as $\text{ReLU}(\bm{\sigma} - \mu\sigma_{K+1})$. Here, $\sigma_{K+1}$ represents the $(K+1)$-th largest singular value of $\mathbf{X}$, while $\mu$ is a trainable parameter that controls the strength of the thresholding, with its value being constrained between 0 and 1 using a sigmoid activation function. This means that we would keep the singular vectors corresponding to the $K$ largest singular values, while we learn how much information to discard through updating $\mu$ using backpropagation. An interesting observation was made in our simulation that the learned $\mu^{(l)}$ generally increases as $l$ increases, which means that the algorithm may be looking to avoid discarding the information at once. Instead, it imposes a gradually stricter model order selection as the data travel down the network. % However, as we are reducing all singular values with the same amount $\mu \sigma_{K+1}$, the order of the magnitude of the singular values remains unchanged and hence the subspace swap event will still occur.

By cascading the iterations, we form a deep unfolded neural network that effectively denoise the sum of exponentials before using Prony's method to reconstruct FRI signals. The detailed algorithm and the corresponding block diagram are shown in \cref{fig:unfoldedwirtinger}. Practically, each iteration layer has its own set of parameters. We initialise the weight matrices across all layers in a way such that it performs exactly as the normal PWGD. For all the simulations in this paper, we adopted a common choice of the constants $\delta_1=\delta_2=0.9999$, initialised $\mu$ as $0.25$, and used $L=5$ unfolded layers.

\subsubsection{Loss Function}

For the loss function, we wish to find a denoised matrix $\hat{\mathbf{S}}$ that best annihilates the ground truth annihilating filter $\mathbf{h}$ which contains the information of the ground truth locations, i.e. minimising $\left\|\hat{\mathbf{S}}\mathbf{h}\right\|_2^2$.

However, similar to Prony's method, we also need to eliminate the trivial solution $\hat{\mathbf{S}}=\mathbf{0}$. This was addressed in the zero eigenvalue-based loss proposed in \cite{Dang2018}, where they added a regularisation term to maximise the projection of $\hat{\mathbf{S}}$ onto the orthogonal complement of $\mathbf{h}$, given by the Frobenius norm of $\overline{\hat{\mathbf{S}}}=\hat{\mathbf{S}}(\textbf{I}-\mathbf{h}\mathbf{h}^H)$. The norm is then put in the exponent so that the regularisation term is bounded by $[0,1]$ for numerical stability. The overall loss function can be expressed by
\begin{align}
    \mathcal{L}(\mathbf{\hat{S}}) = \left\|\hat{\mathbf{S}}\mathbf{h}\right\|^2_2+\alpha e^{-\beta\left\|\overline{\hat{\mathbf{S}}}\right\|^2_F},
\end{align}
where $\alpha$ and $\beta$ are two constants that controls the strength of regularisation, which are respectively set to be $10$ and $0.005$ in the simulation. The learning rate of the unfolded network is set to $2\times10^{-4}$ and each network is trained for 500 epochs. Backpropagation with Adam optimiser \cite{Kingma2015} is used for learning the model.

\subsection{FRI Encoder-Decoder Network (FRIED-Net)}
\begin{figure}[t]
\vspace*{-0.1cm}
    \centering
    \input{figures/autoencoder_structure.tex}
    \caption{The encoder network maps the input noisy samples $\{\tilde{y}[n]\}_{n=0}^{N-1}$ to the estimated locations of the Diracs $\{\hat{t}_k\}_{k=0}^{K-1}$. Depending on the information we have, the amplitudes $\{\hat{a}_k\}_{k=0}^{K-1}$ can either be the ground truth amplitudes or be directly estimated using least squares method fitting $\left\{\hat{\varphi}\left(\hat{t}_k/T-n\right)\right\}_{n=0}^{N-1}$ to $\{\tilde{y}[n]\}_{n=0}^{N-1}$. Given the estimated locations and amplitudes, the decoder, which can either be fixed using knowledge of the sampling kernel $\varphi(t)$ or learned using backpropagation, resynthesises the noiseless samples as $\hat{y}[n]=\sum_{k=0}^{K-1} \hat{a}_k \hat{\varphi}\left(\frac{\hat{t}_k}{T}-n\right)$. }
    \label{fig:autoencoderstructure}
\end{figure}
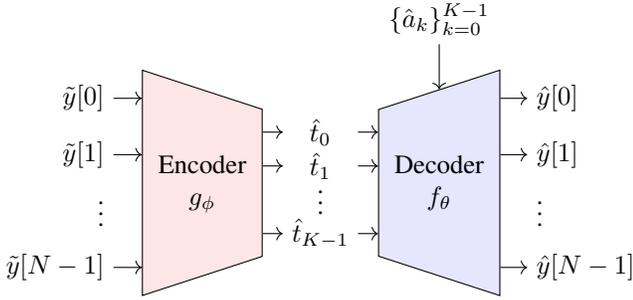

While the deep unfolding approach provides a concrete connection between the iterative denoising algorithm and deep neural networks, it has to be followed by applying the subspace-based Prony's method. Hence, we explore an alternative possibility of bypassing the use of subspace estimation by incorporating knowledge of the FRI acquisition process directly into our network architecture. Since FRI signals are defined by a small number of parameters, we build an autoencoder-like model, named FRI encoder-decoder network (FRIED-Net), by treating the free parameters $\left\{t_k\right\}_{k=0}^{K-1}$ as the latent variables. \cref{fig:autoencoderstructure} outlines our proposed model, with the encoder inferring the estimated locations of the Diracs from the input noisy samples while the decoder resynthesises the noiseless samples from the estimated locations and amplitudes. Depending on the information we have about the sampling kernel $\varphi(t)$, we can opt to fix the decoder or to learn it using backpropagation. We start by describing each of their design and rationale, before delving into the network architecture and learning strategies.

\subsubsection{Encoder Network Design and Architecture}

The encoder network $g_\phi(\cdot): \mathbb{R}^{N} \rightarrow \mathbb{R}^{K}$ infers the locations of the Diracs $\hat{t}_k$ directly from the noisy samples $\tilde{y}[n]$, i.e. $\hat{t}_k=g_\phi\left(\tilde{y}[n]\right)$. Here, the encoder network infers only the locations since the problem of solving the locations is non-linear while estimating the amplitudes is linear, as explained in \cref{sect:classicalfri}. Given the locations, the amplitudes of the Diracs can be directly estimated using a least squares method fitting $\left\{\hat{\varphi}\left(\hat{t}_k/T-n\right)\right\}_{n=0}^{N-1}$ to $\{\tilde{y}[n]\}_{n=0}^{N-1}$.

\begin{figure}[t]
    \vspace*{-0.25cm}
    \centering
    \includegraphics[width=0.9\linewidth]{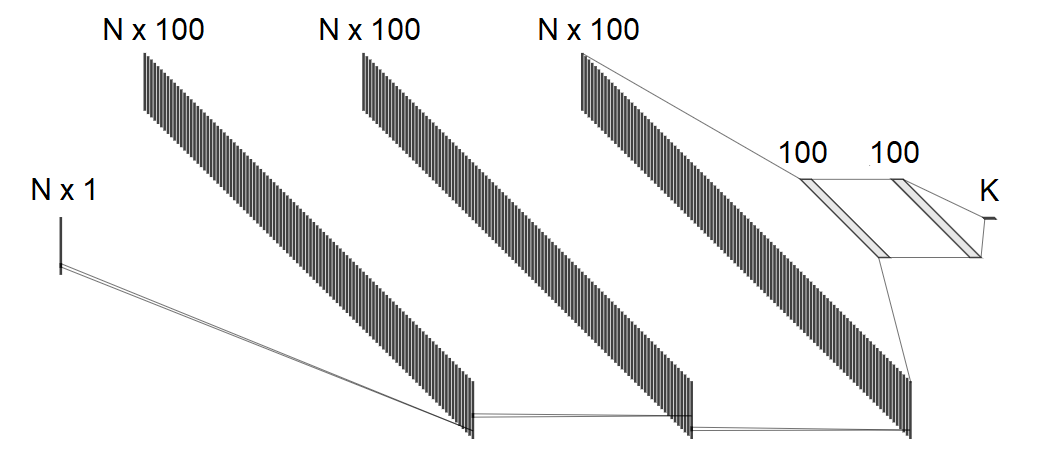}
    \caption{Encoder network architecture to perform inference from the observed noisy samples $\{\tilde{y}[n]\}_{n=0}^{N-1}$ to the locations of Diracs $\{\hat{t}_k\}_{k=0}^{K-1}$.}
    \label{fig:encoderarchitecture}
\end{figure}

As shown in \cref{fig:encoderarchitecture}, our architecture consists of 3 convolutional layers followed by 3 fully connected (FC) layers of sizes $100, 100, K$ respectively. Each of the convolutional layers has 100 filters of size 3. Rectified linear unit (ReLU) is used as the activation function between each two layers. This follows our previous work in \cite{Leung2020} as it empirically provided the best results.

\subsubsection{Decoder Network Design and Architecture}
\begin{figure}[t]
    \vspace*{-0.5cm}
    \centering
    \includegraphics[width=0.95\linewidth]{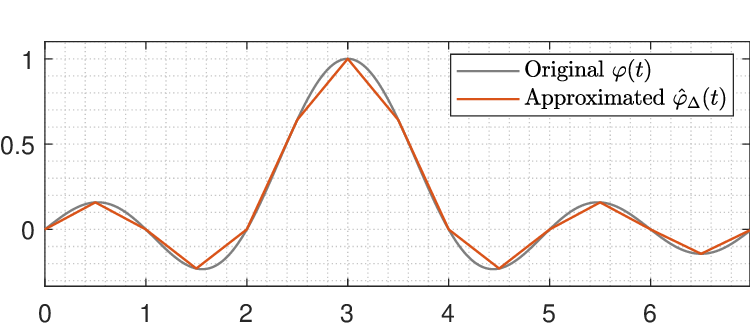}
    \caption{A comparison of an arbitrary sampling kernel $\varphi(t)$ and its corresponding piecewise linear approximation $\hat{\varphi}_\Delta(t)$ using ReLU networks with a uniform step size of $\Delta=1/2$.}
    \label{fig:interpolationexample}
\end{figure}

The decoder network $f_\theta(\cdot): \mathbb{R}^{K} \rightarrow \mathbb{R}^{N}$ aims to transform the estimated locations $\hat{t}_k$ back to the denoised samples $\hat{y}[n]$ using fully connected networks and ReLU, i.e. $\hat{y}[n]=f_\theta(\hat{t}_k)=f_\theta\left(g_\phi\left(\tilde{y}[n]\right)\right)$. This resynthesis problem can be described by 
\begin{align}
    \hat{y}[n]=\sum_{k=0}^{K-1} \hat{a}_k \hat{\varphi} \left(\frac{\hat{t}_k}{T}-n\right), \label{eq:estimatedsamples}
\end{align}
where $\hat{a}_k$ are the estimated amplitudes. Here $\hat{\varphi}(t)$ can be the ground truth sampling kernel $\varphi(t)$ when it is known, or the learned sampling kernel otherwise.

While parameters such as $\hat{a}_k$, $T$, $n$ can be easily modelled as the weights and biases in a fully connected neural network, expressing the sampling kernel using a network and possibly learning it requires a specific design. Works such as \cite{Hornik1989, Leshno1993,Shaham2018} have suggested the capability of ReLU networks as universal approximator of any arbitrary function of compact support. Hence, the decoder can follow the same framework and be used as an approximator of the sampling kernel. Theoretically, the approximation framework in \cite{Hornik1989,Leshno1993} allows to approximate any kernel with an arbitrary and non-uniform resolution with ReLU networks. In this paper, we focus on piecewise linear estimation with a uniform step $\Delta$. The approximated sampling kernel by the ReLU decoder network $\hat{\varphi}_\Delta(t)$ can be expressed as
\begin{align}
    \hat{\varphi}_\Delta(t)=\sum_{i=0}^{I-1} d_i\text{ReLU}(t-i\Delta) \text{,} \label{eq:linearinterpolation}
\end{align}
with the subscript $\Delta$ indicating the piecewise linearity. By utilising $I$ ReLU units, we are effectively dividing the sampling kernel into $I$ linear segments. Therefore, the total number of linear segments is given by $I = L/\Delta$, where $L$ is the support of the kernel. The coefficients $d_i$ are effectively deciding the shape of the estimated kernel $\hat{\varphi}_\Delta(t)$. Depending on whether the ground truth sampling kernel is known, they can be either fixed or learned using backpropagation. For the former case, the coefficients $d_i$ are fixed using the following relationship:
\begin{align}
    d_i = \frac{\varphi\left((i+1)\Delta\right)-\varphi(i\Delta)}{\Delta} - d_{i-1} \quad \text{and} \quad d_0=0. \label{eq:decodercoeff}
\end{align}
\cref{fig:interpolationexample} shows an example of an arbitrary kernel approximated using our decoder with its coefficients $d_i$ fixed according to \eqref{eq:decodercoeff}. We observe that when the step becomes infinitely small $\Delta \to 0$, $\hat{\varphi}_\Delta(t)$ will ultimately converge to the original sampling kernel $\varphi(t)$.

Given the ability to express $\hat{\varphi}_\Delta(\cdot)$ in terms of a ReLU network, we now substitute \eqref{eq:linearinterpolation} into \eqref{eq:estimatedsamples} and express the estimated samples $\{\hat{y}[n]\}_{n=0}^{N-1}$ as 
\begin{align}
    \hat{y}[n] &= \sum_{k=0}^{K-1} \hat{a}_k  \hat{\varphi}_\Delta\left(\frac{\hat{t}_k}{T}-n\right)\\ 
    &= \sum_{k=0}^{K-1} \hat{a}_k \sum_{i=0}^{I-1} d_i\text{ReLU}\left(\frac{\hat{t}_k}{T}-n-i\Delta\right)\text{.} 
\end{align}

\begin{figure}[!t]
\vspace*{-0.15cm}
    \small
    \setlength{\tabcolsep}{2pt}
    \renewcommand{\arraystretch}{2}
    \captionof{table}{Dynamics and Outputs at Each Layer of the Decoder}
    \label{table:decodernetwork}
    \centering
    \begin{tabular}{ccc}
    \toprule
    \noalign{\vspace{-1ex}}
    Layer & Output at Each Layer & \# of Outputs \\ \midrule
    Input & $\left\{\hat{t}_k\right\}_{k=0}^{K-1}$ & $K$ \\
    FC1 & $\left\{\frac{\hat{t}_k}{T} - n\right\}_{k=0,n=0}^{K-1,N-1}$ & $KN$ \\
    FC2+ReLU & $\left\{\text{ReLU}\left(\frac{\hat{t}_k}{T}-n-i\Delta\right)\right\}_{k=0,n=0,i=0}^{K-1,N-1,I-1}$  & $KNI$ \\
    Output & $\left\{\hat{y}[n] = \sum_{k=0}^{K-1} a_k\hat{\varphi}_\Delta\left(\frac{\hat{t}_k}{T}-n\right)\right\}_{n=0}^{N-1}$ & $N$ \\ \bottomrule
    \end{tabular}
    \vspace*{0.3cm}
    \centering
    \input{figures/decoder_architecture.tex}
    \vspace*{-0.6cm}
    \captionof{figure}{An example of decoder architecture for acquiring $N=2$ samples from sampling a stream of $K=1$ Diracs using an approximated sampling kernel $\hat{\varphi}_\Delta(t)$ with $I=3$ linear segments.}
    \label{fig:decoderarchitecture}
    \vspace*{-0.1cm}
\end{figure}

To implement this framework, the decoder consists of 3 fully connected hidden layers of sizes $KN, KNI \text{ and } N$ respectively. The detailed parameters are listed in \cref{table:decodernetwork}. The decoder performs the transformation from the estimated locations produced by the encoder $\{\hat{t}_k \}_{k=0}^{K-1}$ to the estimated samples $\{\hat{y}[n]\}_{n=0}^{N-1}$. An example decoder for $N=2, K=1, I=3$ is also shown in \cref{fig:decoderarchitecture}. Note that in evaluation stage, we would only need the encoder to infer the locations of Diracs from the noisy samples. In this paper, we opt for a high resolution of $\Delta=1/64$, meaning that for every sampling period $T$, we approximate the sampling kernel by 64 linear pieces. 

\subsubsection{Loss Function}
\label{sect:FRIEDlossfunction}
Since we would like the recovered samples to be denoised, the loss function is the squared error between the output estimated samples $\left\{\hat{y}[n] \right\}_{n=0}^{N-1}$ and the noiseless samples $\left\{y[n] \right\}_{n=0}^{N-1}$. Furthermore, we impose a constraint on the bottleneck by including the squared error between the estimated locations $\left\{\hat{t}_k\right\}_{k=0}^{K-1}$ and the ground truth locations $\left\{t_k\right\}_{k=0}^{K-1}$. This is necessary because there exists an ambiguity: when the kernel shifts by an arbitrary amount $\epsilon$, the same set of samples can be obtained when we add a bias $-T\epsilon$ to the locations $t_k$. Using \eqref{eq:estimatedsamples}, it can be shown mathematically as follows:
\begin{align}
    \hat{y}[n] =\sum_{k=0}^{K-1} \hat{a}_k \hat{\varphi} \left(\frac{\hat{t}_k}{T}-n\right) &= \sum_{k=0}^{K-1} \hat{a}_k \hat{\varphi} \left(\frac{\hat{t}_k-T\epsilon}{T}-n+\epsilon\right) \nonumber \\
    &= \sum_{k=0}^{K-1} \hat{a}_k \hat{\varphi}' \left(\frac{\hat{t}'_k}{T}-n\right),
\end{align}
where $\hat{\varphi}'(t) = \hat{\varphi}(t+\epsilon)$ and $\hat{t}'_k = \hat{t}_k-T\epsilon$.

Together, the resultant loss function can be written as
\begin{align}
    \mathcal{L}(\mathbf{\hat{y}},\mathbf{\hat{t}})&=  \sum_{n=0}^{N-1}\left(\hat{y}[n]-y[n]\right)^2
    +\gamma\sum_{k=0}^{K-1}\left(\hat{t}_k-t_k\right)^2\text{,} \label{eq:lossfunction}
\end{align}
where $\gamma$ is a constant which controls the strength of the constraint on the bottleneck. Backpropagation with Adam optimiser \cite{Kingma2015} is used for learning the model. The learning rate of the encoder and the decoder are set to be $10^{-4}$, $10^{-5}$ respectively. The value $\gamma$ is set to 1 for \cref{sect:knownkernel} and 100 for \cref{sect:unknownkernel} and \cref{sect:calciumresults}. 

\subsubsection{Training Strategies}
\label{sect:trainingstrat}

Since FRIED-Net can be used either when the sampling kernel $\varphi(t)$ is known or when it is unknown, we deploy two different training strategies according to the situation. 

\paragraph{Known sampling kernel $\varphi(t)$}

As classical FRI algorithms require knowledge of the sampling kernel $\varphi(t)$, we first consider that scenario for our proposed FRIED-Net. We also further assume that we have the information of the noiseless samples $\left\{y[n] \right\}_{n=0}^{N-1}$ and thus the amplitudes of the pulses $\left\{a_k \right\}_{k=0}^{K-1}$ in the \emph{training} data. Given this information, we then fix the parameters of the decoder network using the relationship in \eqref{eq:decodercoeff}. Note that during testing, since only the encoder is used to estimate the locations from the noisy samples, the ground truth amplitudes of the test data are not required.

For training, we adopt a warm start approach, which means that the encoder is first initialised using the trained direct inference encoder network. This provides an initial estimation of the reconstructed FRI parameters. We then incorporate the decoder and the encoder network is trained for 150 further epochs, during which the weights and the biases of the decoder are frozen as the decoder is modelled from approximating the true sampling kernel $\varphi(t)$. This \emph{fixed decoder} provides an implicit and accurate regularisation on the estimated pulse locations of the encoder network and therefore, fine-tunes the learning of the encoder network.

\paragraph{Unknown sampling kernel $\varphi(t)$}
On the other hand, we would like to overcome the constraint of the classical FRI algorithms and reconstruct an FRI signal without knowledge of sampling kernel. Hence, we propose to learn the coefficients $\mathbf{d}$ of the decoder through backpropagation, which effectively translates to estimating the sampling kernel $\hat{\varphi}_\Delta(t)$. Contrary to previous assumptions that the ground truth samples $\left\{y[n] \right\}_{n=0}^{N-1}$ and the amplitudes of the pulses $\left\{a_k \right\}_{k=0}^{K-1}$ are known, they are now replaced by the noisy samples $\{\tilde{y}[n]\}_{n=0}^{N-1}$ and the amplitudes estimated by least squares fitting $\left\{\hat{\varphi}_\Delta\left(\hat{t}_k/T-n\right)\right\}_{n=0}^{N-1}$ to $\{\tilde{y}[n]\}_{n=0}^{N-1}$. 

Another change regards the decoder coefficients $\mathbf{d}$, caused by the lack of knowledge of the ground truth sampling kernel. In the previous simulation, as the ground truth sampling kernel is known, we fixed them using the relationship in \eqref{eq:decodercoeff}. In this scenario, we initialised the coefficients to $\mathbf{d} \sim \mathcal{U}(-0.01,0.01)$ to ensure that the initial estimated kernel would be a non-zero signal, and made them learnable via backpropagation.

However, now there exists ambiguity. From \eqref{eq:samples}, we observe that the samples are the sum of the products of shifted versions of the sampling kernel and the amplitudes of the pulses. When the sampling kernel is unknown, assuming the true amplitudes and kernel are $\left\{a_k \right\}_{k=0}^{K-1}$ and $\varphi(t)$ respectively, then $\left\{\zeta a_k \right\}_{k=0}^{K-1}$ and $\varphi(t)/\zeta$ for any real factor $\zeta$ are also valid choices of the amplitudes and sampling kernel that synthesise identical samples $\{y[n]\}_{n=0}^{N-1}$. To avoid this ambiguity in the kernel while training FRIED-Net, we fix the peak value of the estimated kernel $\hat{\varphi}_\Delta(t)$ to be 1 by normalising the coefficients $\mathbf{d}$ after each epoch. This is done by 
\begin{align}
\mathbf{d}_{\text{norm}} = 
    \begin{cases}
    \frac{\mathbf{d}}{\max_t (\hat{\varphi}_\Delta)} & \text{if } \abs{\max_t (\hat{\varphi}_\Delta)} \geq \abs{\min_t (\hat{\varphi}_\Delta)} \\
    \frac{\mathbf{d}}{\min_t (\hat{\varphi}_\Delta)} &  \text{otherwise.}
    \end{cases}
\end{align}

We also modify slightly the way in which we train the network since we now have to learn also the decoder. Previously, we initialised the encoder network with our trained model using a direct inference method, incorporated the fixed decoder and trained the encoder for 150 epochs. Here, we keep the warm start approach but then train the decoder for 150 epochs with the parameters of the encoder frozen, before training the entire network jointly for another 150 epochs. Effectively, the initialisation from direct inference provides a coarse estimate of the locations such that the decoder can reference and hence learn a rough estimate of the sampling kernel. Eventually, we train the entire network in order to refine the estimations of both the locations and the sampling kernel. 

\section{Simulation}
\label{sect:simulation}

In this section, we present simulation results of our proposed algorithms in different scenarios of reconstructing a periodic stream of $K$ Diracs with $t_k \in [-0.5, 0.5)$ and $a_k \in \mathbb{R}^+$ under noisy conditions. 

To evaluate the performance, the samples $\left\{y[n] \right\}_{n=0}^{N-1}$ are corrupted with additive white Gaussian noise at different PSNR $\in [-5, 70]$ dB with a step of 5 dB. Here, PSNR is defined by the ratio between the maximum amplitude of each signal and the standard deviation of Gaussian noise, which is expressed as:
\begin{align}
\text{PSNR} = 20 \log_{10} \left(\frac{\max_{k} a_k}{\sigma_\varepsilon}\right).
\end{align}
The metric we use is the standard deviation of the retrieved location of Diracs, defined as: 
\begin{align}
SD_k=\sqrt{\frac{\sum_{j=0}^{J-1}\left(\hat{t}_k^{(j)}-t_k\right)^2}{J}}\text{,} \label{eq:sd}
\end{align}
where $\hat{t}_k^{(j)}$ and $J$ are the $j$-th estimation and the number of realisations respectively.

In all the simulations, the number of samples and signal period are set to $N=21$ and $\tau = 1$ respectively. An individual network is trained for each PSNR using PyTorch \cite{Paszke2019}.\footnote{For reproducibility, our source code is available at \url{https://github.com/vchleung/LearningBasedFRI}.} The number of training data for Deep Unfolded PWGD and FRIED-Net is $10^6$. We set $t_k \sim \mathcal{U}(-0.5,0.5)$ and $a_k \sim \mathcal{U}(0.5,10)$ for $k=0,1$, where $\mathcal{U}(a,b)$ denotes uniform distribution between $a$ and $b$, and generate both the training data and test data using the same sampling kernel $\varphi(t)$.

\subsection{Reconstruction with Known Sampling Kernel \texorpdfstring{$\varphi(t)$}{phi(t)}}
\label{sect:knownkernel}

We start by applying our proposed learning-based approaches and classical FRI techniques when the sampling kernel is known, such that we can compare the performance in terms of the breakdown PSNR \cite{Wei2015}. We choose the sampling kernel $\varphi(t)$ to be an exponential reproducing kernel of maximum order and minimum-support (eMOMS) \cite{Uriguen2013} that can reproduce $P+1=N$ exponentials with $\omega_0 = \frac{-P\pi}{P+1}$ and $\lambda =\frac{2\pi}{P+1}$. 

\subsubsection{\texorpdfstring{$K=2$}{K=2}}

We first focus on a simple case of having $N=21$ samples, synthesised from a stream of $K=2$ Diracs with equal amplitudes $a_0 = a_1 \sim \mathcal{U}(0.5,10)$ in the evaluation stage. This allows to compare our results with the breakdown PSNR shown in \cref{fig:breakdownPSNR}. We fix the first Dirac at $t_0=0.1$ and change $\Delta t_0 \in [10^{-0.5},10^{-3}]$ evenly on a logarithmic scale with a step size of $10^{-0.25}$. Monte Carlo simulations with 10000 realisations are performed for each PSNR-$\Delta t_0$ pair.

\begin{figure*}[t]
\vspace*{-0.5cm}
\centering
\subfloat[Prony's method with Cadzow \\ denoising.]{
\centering
\includegraphics[width=0.23\linewidth]{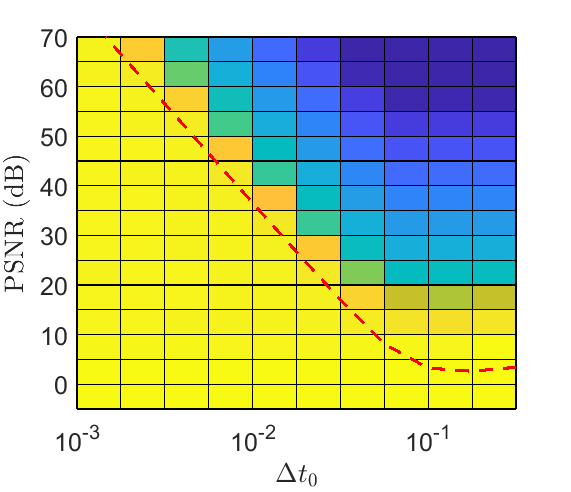}}
\hfil
\subfloat[Prony's method with Deep Unfolded PWGD. \label{fig:eMOMS_unfolding}]{
\centering
\includegraphics[width=0.23\linewidth]{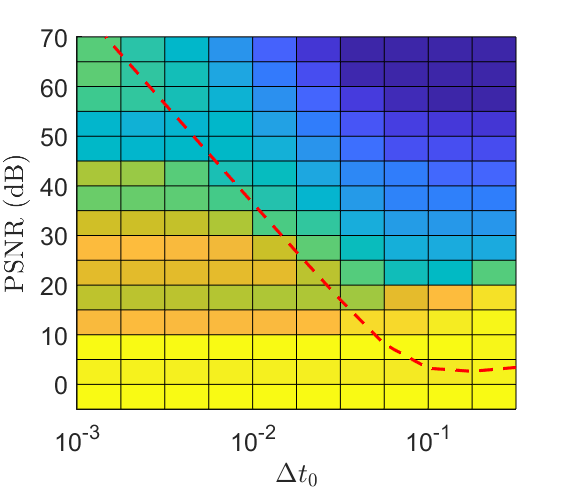}}
\hfil
\subfloat[Direct inference using DNN \cite{Leung2020} (equivalent to training only the encoder of FRIED-Net). \label{fig:eMOMS_DI}]{
\centering
\includegraphics[width=0.23\linewidth]{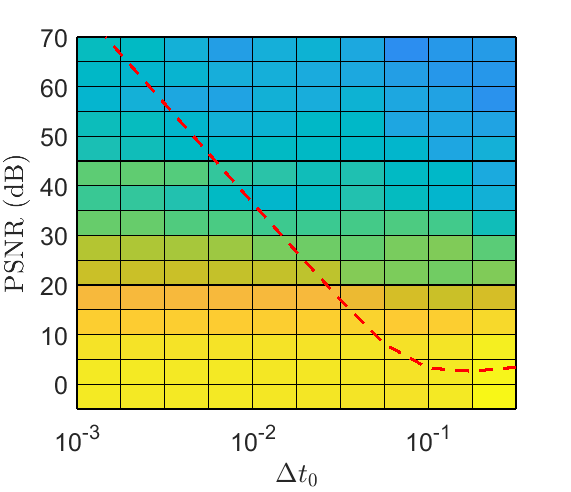}}
\hfil
\subfloat[FRIED-Net. \label{fig:eMOMS_AE}]{
\centering
\includegraphics[width=0.23\linewidth]{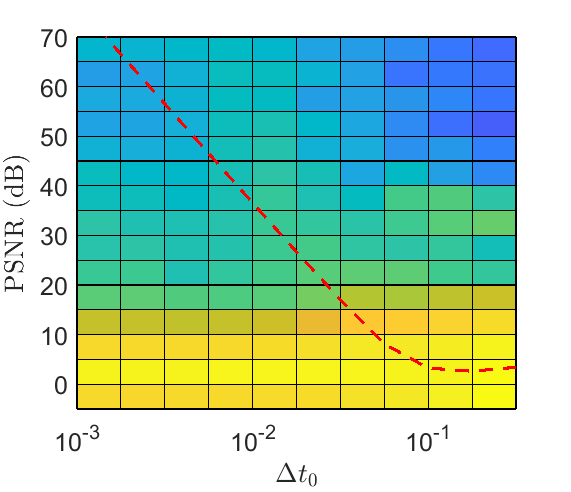}}
\hspace*{-0.4cm}
\includegraphics[width=0.042\linewidth]{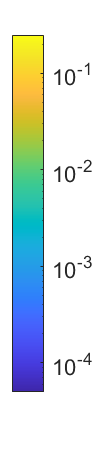}
\\
\vspace*{-0.2cm}
\centering
\subfloat[FRIED-Net fine-tuned by backpropagating {$\sum_{n}\left(\tilde{y}[n]-\hat{y}[n]\right)^2$} for each test datum. \label{fig:eMOMS_backprop}]{
\centering
\includegraphics[width=0.23\linewidth]{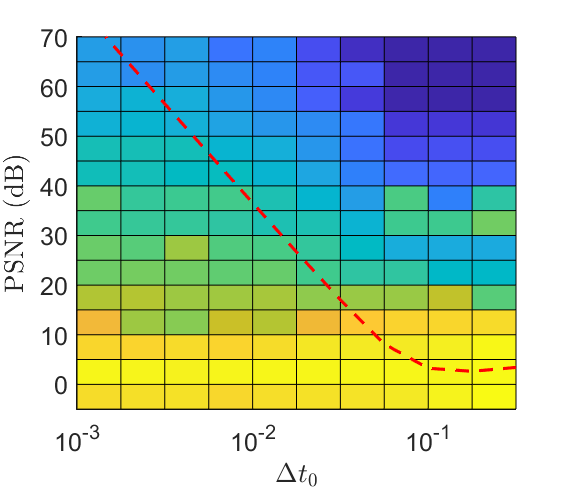}}
\hfil
\subfloat[DeepFreq \cite{Izacard2019}. \label{fig:eMOMS_DeepFreq}]{
\centering
\includegraphics[width=0.23\linewidth]{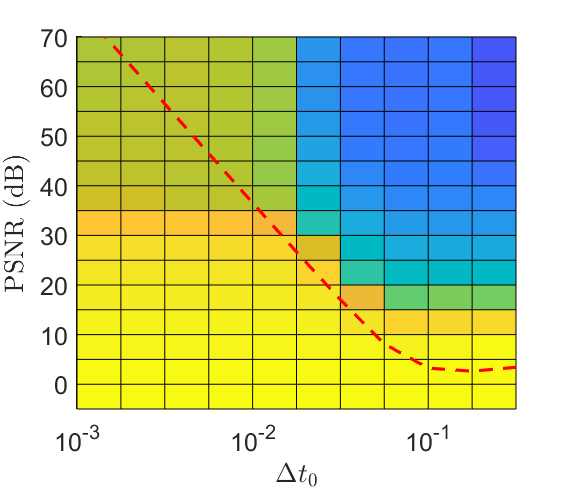}}
\hfil
\subfloat[Direct inference \\(single model for all PSNRs). \label{fig:eMOMS_DI_singlemodel}]{
\centering
\includegraphics[width=0.23\linewidth]{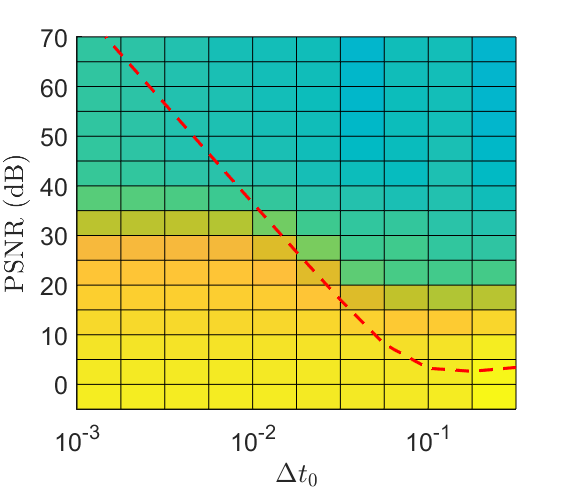}}
\hfil
\subfloat[Fine-tuned FRIED-Net \\(single model for all PSNRs).\label{fig:eMOMS_backpropGD_singlemodel}]{
\centering
\includegraphics[width=0.23\linewidth]{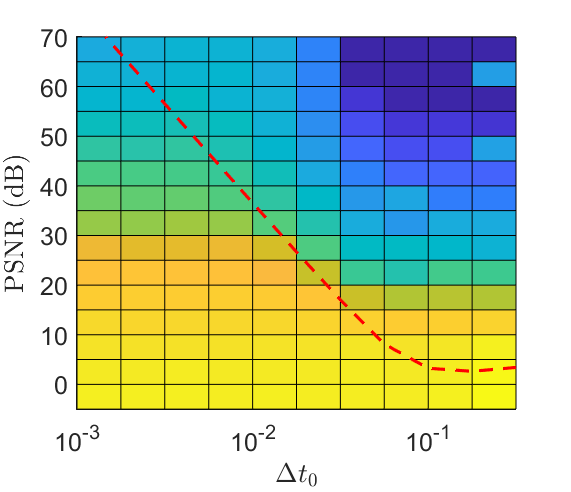}}
\hspace*{-0.4cm}
\includegraphics[width=0.042\linewidth]{figures/colorbar.png}
\caption{Mean standard deviation of the retrieved locations of a stream of Diracs sampled by eMOMS ($P+1=N=21, K=2$) over 10000 realisations at each PSNR-$\Delta t_0$ pair using different methods. The red dashed line refers to the breakdown PSNR calculated using \eqref{eq:breakdownPSNR} \cite{Wei2015}. The reconstruction performance is better when the colour of the grid is darker, indicating a low mean standard deviation, and vice versa.}
\label{fig:eMOMSresults}
\vspace*{-0.2cm}
\end{figure*}

We first compare the reconstruction performance of our previous work in direct inference method (equivalent to training only the encoder of FRIED-Net) \cite{Leung2020}, our proposed FRIED-Net and Deep Unfolded PWGD against the classical subspace-based Prony's method with Cadzow denoising \cite{Prony1795,Cadzow1988} using mean standard deviation. \cref{fig:eMOMSresults} shows the respective results, with the breakdown PSNR plotted in \cref{fig:breakdownPSNR} overlaid as the red dashed line to aid visualisation. As discussed in \cref{sect:breakdownpsnr}, we see that the performance of Prony's method with Cadzow denoising suffers from an abrupt deterioration below the red dashed line. This demonstrates the breakdown in performance due to the inherent subspace swap event in subspace-based approaches. On the other hand, all of our proposed learning-based algorithms maintain consistent performance across different $\Delta t_0$. For instance, when $\Delta t_0=10^{-2}$, Deep Unfolded PWGD and FRIED-Net break down at around PSNR $=$ 15 dB whereas Prony's method with Cadzow denoising breaks down at PSNR $=$ 40 dB. It shows that solving the original FRI reconstruction problem through learning-based approaches enables to recover FRI signals with a higher resolution under strong noise. We also observe that FRIED-Net performs the best amongst the proposed algorithms in the breakdown region, as indicated by the spread of low standard deviation region below the breakdown PSNR curve at around 15 dB in \cref{fig:eMOMS_AE}. This shows that the decoder plays an important role in regularisation and in fine-tuning the estimations.

We next look into the low noise regimes. We observe that Deep Unfolded PWGD maintains very similar reconstruction performance to the classical technique. Nonetheless, FRIED-Net comes with a slight compromise in the top right regions of \cref{fig:eMOMSresults}, which can usually be eased using gradient descent that is based on the squared error $\sum_{n=0}^{N-1}\left(\tilde{y}[n]-\hat{y}[n]\right)^2$ between the noisy samples and the samples resynthesised from the retrieved locations of the trained model and the least squares fitted amplitudes. On the other hand, since the exact calculation of the gradient requires a closed-form expression of the derivative of the sampling kernel, we can instead make use of our FRIED-Net architecture and perform fine-tuning through backpropagation of the sample error per test datum. \cref{fig:eMOMS_backprop} shows that FRIED-Net is able to achieve satisfactory results compared to classical FRI methods in the low noise regimes after fine-tuning.

In addition, we compare our results against DeepFreq \cite{Izacard2019}, a learning-based spectral estimation algorithm which outputs a learned spectral representation of the multisinusoidal signal with local maxima at the position of the estimated frequencies. We again use the same training dataset that contains $10^6$ streams of $K=2$ Diracs with $t_k \sim \mathcal{U}(-0.5,0.5)$ and $a_k \sim \mathcal{U}(0.5,10)$. On the other hand, as DeepFreq considers the noise level to be unknown, we instead add new noise realisations at each epoch during training. For each new noise realisation, the noise level is determined by sampling the PSNR from $\sim \mathcal{U}(0, 70)$ dB. 

\cref{fig:eMOMS_DeepFreq} shows the reconstruction performance of DeepFreq. We can observe that DeepFreq provides only a slight improvement in the regions of interest under the red dashed curve, where classical FRI methods break down, and is outperformed by both our proposed Deep Unfolded PWGD and FRIED-Net. On the other hand, in the low noise regimes, despite a slightly better performance than the untuned FRIED-Net, DeepFreq is unable to match that of classical methods, of our proposed Deep Unfolded PWGD and of fine-tuned FRIED-Net. This is because the frequency estimates are obtained by finding the peaks from the output spectral representation of the DeepFreq network. Therefore, unlike our proposed methods, the locations $t_k$ reconstructed by DeepFreq effectively lies on a grid, with its precision limited by the output size of the network. 

We are then also interested in whether our proposed methods can cope with different noise levels using a single model like DeepFreq. Hence, we follow the training framework of DeepFreq that adds new noise realisations at each epoch, with the PSNR sampled from $\sim \mathcal{U}(0, 70)$ dB, and retrain our proposed networks. \cref{fig:eMOMS_DI_singlemodel} and \cref{fig:eMOMS_backpropGD_singlemodel} shows the single-model performance of the direct inference method and the fine-tuned FRIED-Net. Both of our proposed algorithms are able to maintain a similar behaviour over all PSNR-$\Delta t_0$ pairs and overcome breakdown events, despite a slight drop in overall performance.

\begin{figure}[t]
\vspace*{-0.5cm}
\centering
\subfloat[A single model for all noise \\levels. \label{fig:eMOMS_unfolding_singlemodel}]{
\centering
\hspace*{-0.2cm}
\includegraphics[width=0.46\linewidth]{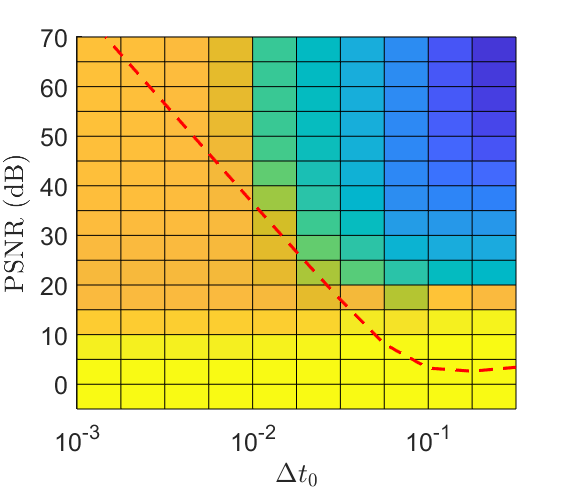}}
\hfil
\subfloat[Models trained for every 20 dB of PSNR (three models in total).\label{fig:eMOMS_unfolding_20dB}]{
\centering
\includegraphics[width=0.46\linewidth]{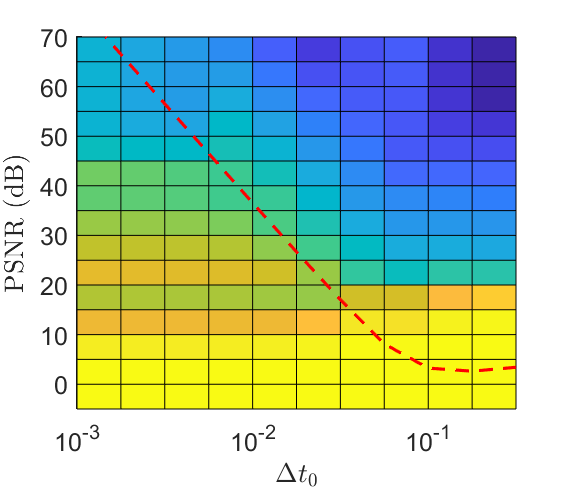}}
\hspace*{-0.4cm}
\includegraphics[width=0.084\linewidth]{figures/colorbar.png}
\caption{Comparison of the mean standard deviation of the retrieved locations of a stream of Diracs sampled by eMOMS ($P+1=N=21, K=2$) using Deep Unfolded PWGD between a single-model and a slightly relaxed multi-model setup.}
\label{fig:eMOMSUnfoldingSingleVsMulti}
\vspace*{-0.2cm}
\end{figure}

Nonetheless, Deep Unfolded PWGD suffers from a notable drop in terms of reconstruction performance within the breakdown region, albeit still performing better than Cadzow denoising, as seen in \cref{fig:eMOMS_unfolding_singlemodel} below the red dashed line. Therefore, we try to relax this training framework slightly by training each model for a closer range of PSNRs and utilising multiple models. \cref{fig:eMOMSUnfoldingSingleVsMulti} shows the performance where we train a model for every PSNR interval of 20 dB, which is effectively an intermediate case between \cref{fig:eMOMS_unfolding} (where we train an individual model for each PSNR) and \cref{fig:eMOMS_unfolding_singlemodel} (where a single model is used for all noise levels). We can see that this intermediate training framework allows Deep Unfolded PWGD to closely match the performance shown in \cref{fig:eMOMS_unfolding}, especially in the regions of interest where the classical FRI methods break down. 

\subsubsection{\texorpdfstring{$K=10$}{K=10}}
\label{sect:10Diracs}

We then move onto the case of reconstructing more pulses. We consider the case of critical sampling, where we are reconstructing $K=10$ Diracs from $N=21$ samples. In evaluation, we assume a case of  Diracs with equal amplitudes $a_k \sim \mathcal{U}(0.5,10)$. For the locations, they are distributed uniformly across the entire timescale, i.e. $t_k\sim \mathcal{U}(-0.5,0.5)$. For each PSNR, Monte Carlo simulations with 10000 realisations are performed.

\begin{figure}[t]
\centering
\subfloat[Mean standard deviation]{
\centering
\includegraphics[width=0.9\linewidth]{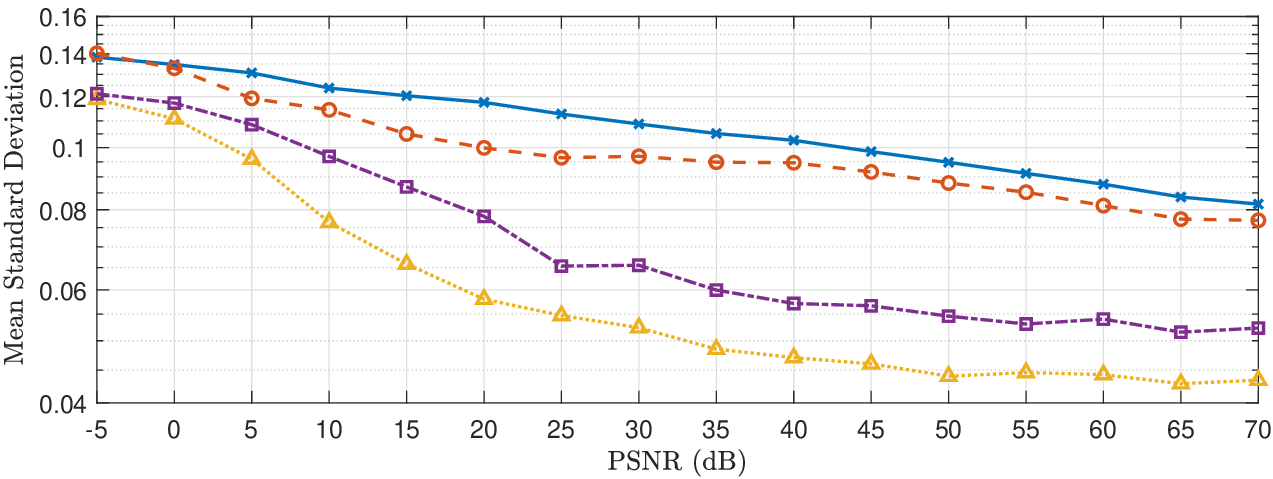}}
\\
\vspace*{-0.3cm}
\centering
\subfloat[Median standard deviation]{
\centering
\includegraphics[width=0.9\linewidth]{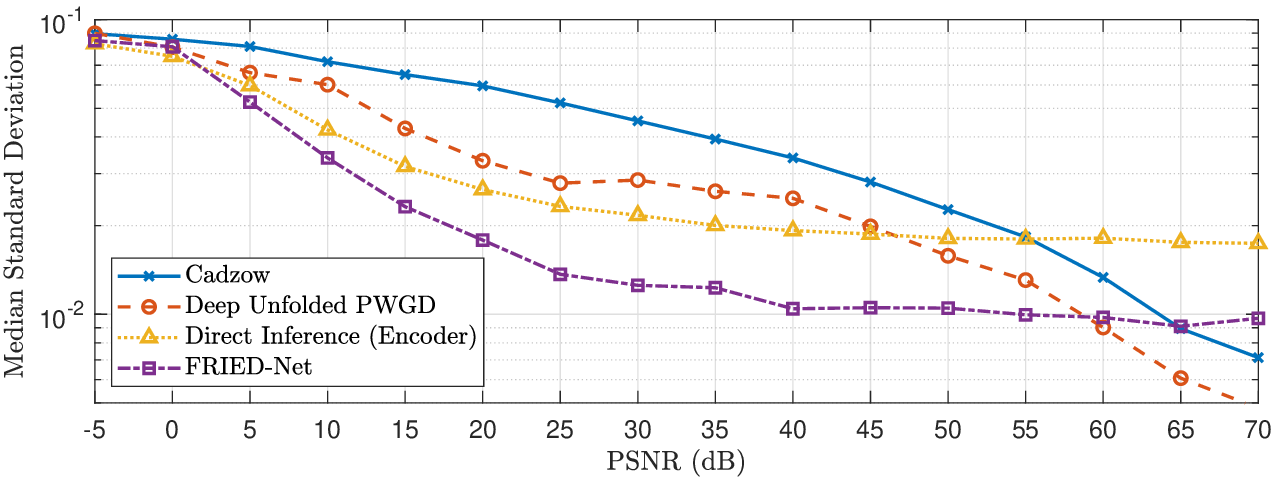}}
\caption{Mean and median standard deviation of the retrieved locations of a stream of Diracs sampled by eMOMS ($P+1=N=21, K=10$) over 10000 realisations at different PSNR using different methods.}
\label{fig:eMOMS_K10}
\end{figure}

\begin{figure}[t]
\vspace*{-0.5cm}
\centering
\subfloat[Prony's method with Cadzow denoising.]{
\centering
\includegraphics[width=0.95\linewidth]{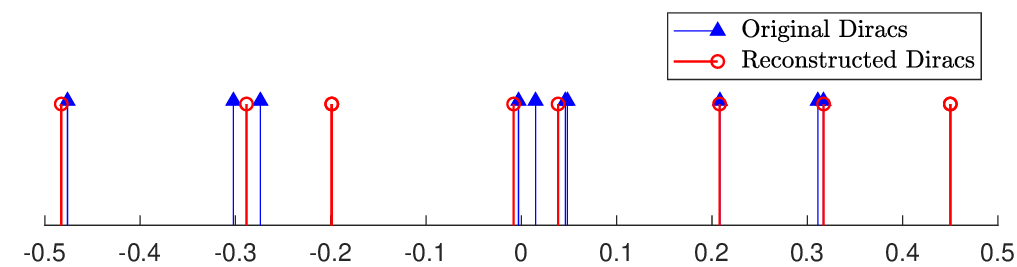}}
\\
\vspace*{-0.2cm}
\subfloat[Prony's method with Deep Unfolded PWGD.]{
\centering
\includegraphics[width=0.95\linewidth]{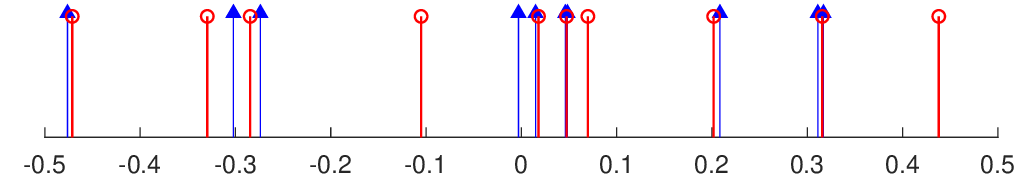}}
\\
\vspace*{-0.2cm}
\subfloat[Direct inference using DNN (Encoder of FRIED-Net).]{
\centering
\includegraphics[width=0.95\linewidth]{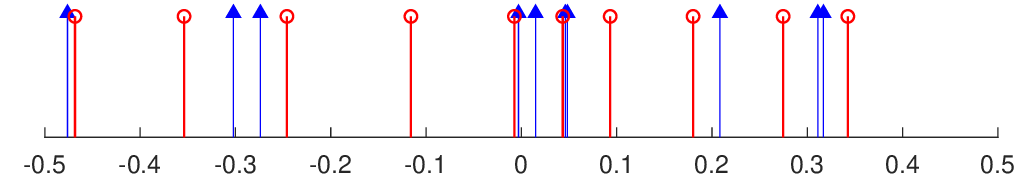}}
\\
\vspace*{-0.2cm}
\subfloat[FRIED-Net]{
\centering
\includegraphics[width=0.95\linewidth]{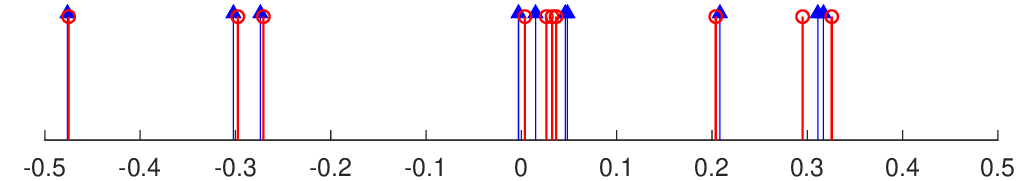}}
\caption{An example of recovered locations of a stream of Diracs sampled by eMOMS ($P+1=N=21, K=10$) at PSNR $=20$ dB using different methods.}
\label{fig:eMOMS_K10example}
\vspace*{-0.1cm}
\end{figure}

\cref{fig:eMOMS_K10} shows the reconstruction performance using different approaches. We use both the mean and median standard deviation across all Diracs, since any missed or falsely detected Diracs may now have a huge impact on the mean standard deviation due to the problem of aligning the order of the reconstructed Diracs and the ground truth. Both plots show that our proposed techniques outperform the classical subspace-based methods. While the encoder of FRIED-Net provides better mean standard deviation, the full FRIED-Net performs better in terms of median standard deviation in high noise levels. This can be further analysed using a representative example at PSNR $=20$ dB in \cref{fig:eMOMS_K10example}. We observe that Prony's method with Cadzow denoising is missing two Diracs in its estimation, despite reconstructing the remaining Diracs fairly precisely. This leads to a misalignment of Diracs and hence a huge penalty especially on the mean standard deviation. Comparatively, Deep Unfolded PWGD is able to improve the precision of the estimation and potentially recover the missing Diracs in the classical approach. In contrast, the acquisition model-inspired FRIED-Net behaves differently as it bypasses the subspace estimation. We can see that the encoder of FRIED-Net provides just a rough estimate of the locations, yielding a lower mean standard deviation. The incorporation of the decoder allows the network to estimate much more precisely, resulting in lower median standard deviation.


\subsection{Reconstruction with Unknown Sampling Kernel \texorpdfstring{$\varphi(t)$}{phi(t)}}
\label{sect:unknownkernel}

Previously, we have shown that both of our proposed learning-based systems can overcome the breakdown PSNR when the sampling kernel is known. In this section, we would like to relax the constraint and reconstruct the signal under the assumptions that neither the sampling kernel nor the noiseless samples are known. The former is motivated by the fact that the sampling kernel $\varphi(t)$ has to be known to find the coefficients $c_{m,n}$ in \eqref{eq:exponentialreproducing} in classical FRI techniques, while the latter is due to the limited information we usually possess in real-world reconstruction problems. Here, we show that our proposed FRIED-Net is capable of reconstructing FRI signals while only possessing the information of the ground truth locations of the training data $\left\{t_k \right\}_{k=0}^{K-1}$ and the noisy discrete samples $\left\{\tilde{y}[n] \right\}_{n=0}^{N-1}$. This is also something that happens in certain neuroscience settings as we will show in \Cref{sect:application}. We also show that although the sampling kernel $\varphi(t)$ is unknown, we are still able to estimate it via learning using the fact that the training data are generated using the same $\varphi(t)$.

\subsubsection{\texorpdfstring{$K=2$}{K=2}, eMOMS}
\label{sect:eMOMSunknown}

\begin{figure}[t]
\vspace*{-0.2cm}
\captionsetup[subfloat]{captionskip=0cm}
\centering
\subfloat[PSNR $=70$ dB]{
\centering
\includegraphics[width=0.9\linewidth]{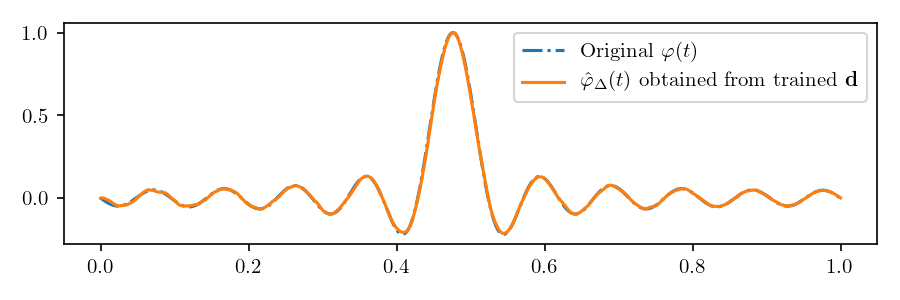}
}
\vspace*{-0.25cm}
\subfloat[PSNR $=10$ dB]{
\centering
\includegraphics[width=0.9\linewidth]{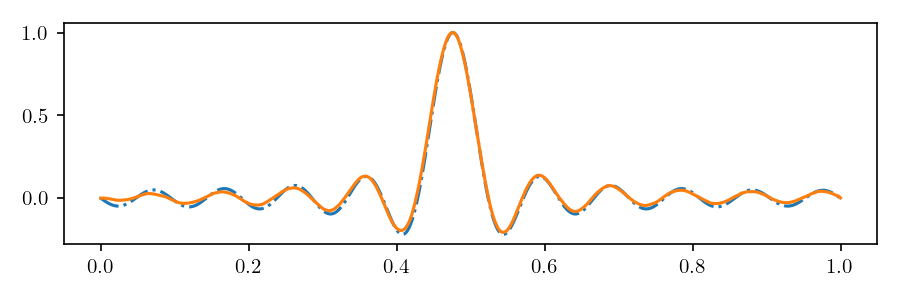}
}
\vspace*{-0.25cm}
\subfloat[PSNR $=0$ dB]{
\centering
\includegraphics[width=0.9\linewidth]{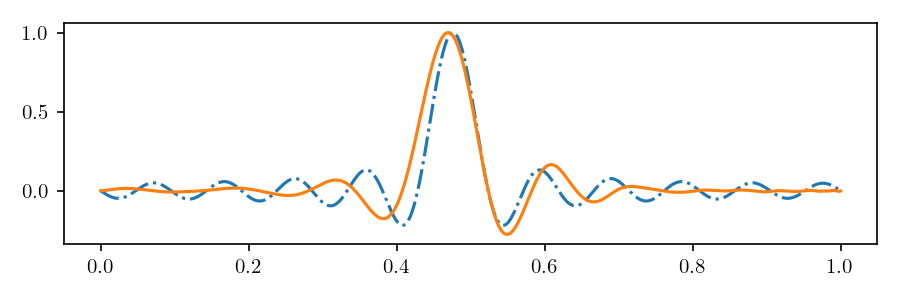}
}
\caption{The estimated kernel $\hat{\varphi}_\Delta(t)$ obtained from the learned coefficients $\mathbf{d}$ of FRIED-Net compared with the ground truth eMOMS $\varphi(t)$.}
\label{fig:eMOMS_estimated}
\vspace*{-0.3cm}
\subfloat[Performance when the sampling kernel is known (from \cref{fig:eMOMS_AE}).]{
\centering
\hspace*{-0.2cm}
\includegraphics[width=0.46\linewidth]{figures/eMOMS_AE.eps}}
\hfil
\subfloat[Performance when the sampling kernel is unknown.]{
\centering
\includegraphics[width=0.46\linewidth]{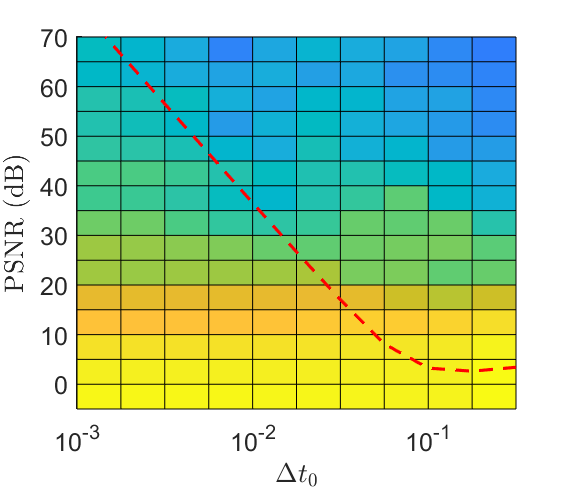}}
\hspace*{-0.4cm}
\includegraphics[width=0.084\linewidth]{figures/colorbar.png}
\caption{Comparison of the mean standard deviation of the retrieved locations of a stream of Diracs sampled by eMOMS ($P+1=N=21, K=2$) using FRIED-Net when the sampling kernel is known or unknown. The red dashed line refers to the breakdown PSNR calculated using \eqref{eq:breakdownPSNR} \cite{Wei2015}.}
\label{fig:eMOMSresultsunknown}
\vspace*{-0.2cm}
\end{figure}

Here we repeat the simulation of \Cref{sect:knownkernel} to compare the performance when the sampling kernel is known or not. Therefore, the sampling kernel $\varphi(t)$ we use to generate both training and test data is again chosen to be an eMOMS. We focus on a simple case of having two Diracs with equal amplitudes $a_0 = a_1 \sim \mathcal{U}(0.5,10)$. Similarly, we fix the first Dirac at $t_0=0.1$ and change $\Delta t_0 \in [10^{-0.5},10^{-3}]$ evenly on a logarithmic scale with a step of $10^{-0.25}$. Monte Carlo simulations with 10000 realisations are performed for each PSNR-$\Delta t_0$ pair.

We begin with visualising the estimated sampling kernels $\hat{\varphi}_\Delta(t)$ in \cref{fig:eMOMS_estimated}. We observe that the network is capable of learning it. The estimated $\hat{\varphi}_\Delta(t)$ matches the original sampling kernel for PSNR up to 10 dB. Nonetheless, when PSNR $=0$ dB, the network is only able to capture the main peak. This shows that theoretically, removing the information of the sampling kernel has a limited impact on the performance of FRIED-Net, apart from extremely noisy conditions.

Next, we compare the performance of the learning-based approach with known kernel (fixed decoder) and unknown kernel (learned decoder). \cref{fig:eMOMSresultsunknown} shows that our proposed FRIED-Net overcomes the breakdown PSNR, as highlighted by the red dashed line, in both circumstances, indicated by the spread of the low standard deviation (blue) region across the red line. On the other hand, we also see the slight overall performance drop compared with the previous simulation when shape of the kernel is known, despite the network learning the sampling kernel well and close to the ground truth. This performance drop is very likely due to the fact that we have also removed the information of the noiseless samples and hence the ground truth information of the amplitudes during training.

\subsubsection{\texorpdfstring{$K=2$}{K=2}, E-Spline}

\begin{figure}[!b]
\vspace*{-0.2cm}
\captionsetup[subfloat]{captionskip=0cm}
\centering
\subfloat[PSNR $=70$ dB]{
\centering
\includegraphics[width=0.9\linewidth]{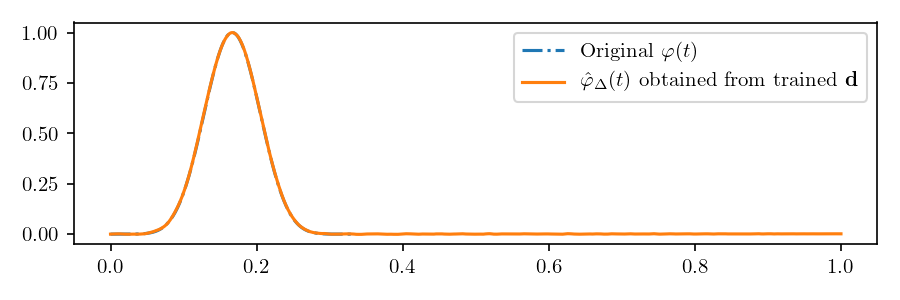}
}
\vspace*{-0.25cm}
\subfloat[PSNR $=10$ dB]{
\centering
\includegraphics[width=0.9\linewidth]{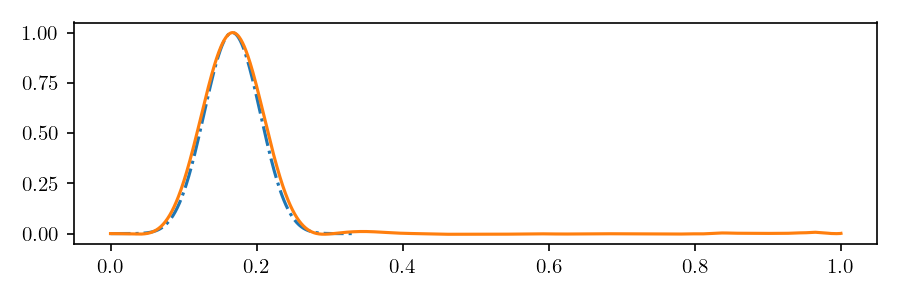}
}
\vspace*{-0.25cm}
\subfloat[PSNR $=0$ dB]{
\centering
\includegraphics[width=0.9\linewidth]{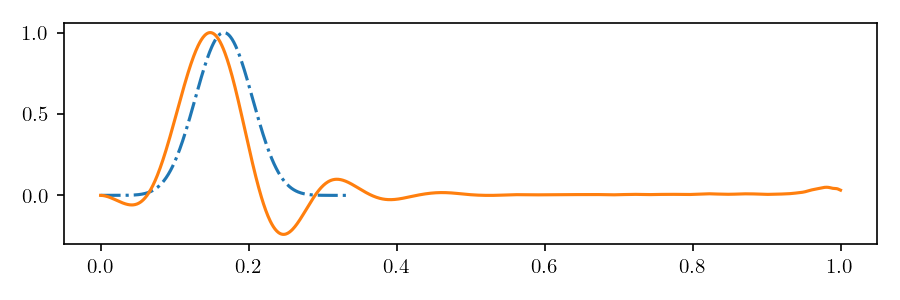}
}
\caption{The estimated kernel $\hat{\varphi}_\Delta(t)$ obtained from the learned coefficients $\mathbf{d}$ compared with the ground truth E-Spline $\varphi(t)$.}
\label{fig:ESpline_estimated}
\end{figure}
\begin{figure}[t]
\vspace*{-0.5cm}
\centering
\subfloat[Prony's method with Cadzow \\denoising. (Known sampling \\kernel)]{
\centering
\hspace*{-0.2cm}
\includegraphics[width=0.46\linewidth]{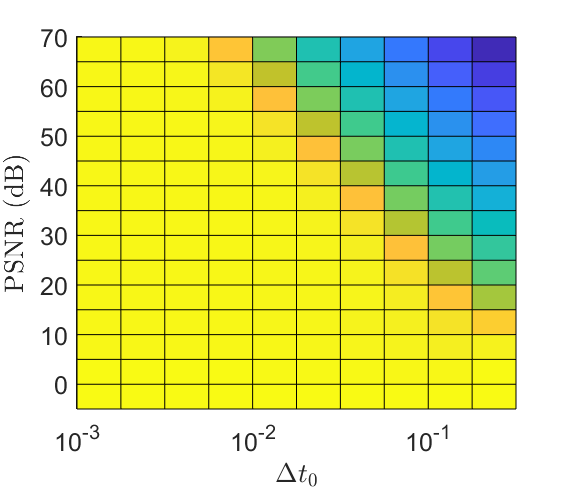}}
\hfil
\subfloat[FRIED-Net without knowledge of the sampling kernel.]{
\centering
\includegraphics[width=0.46\linewidth]{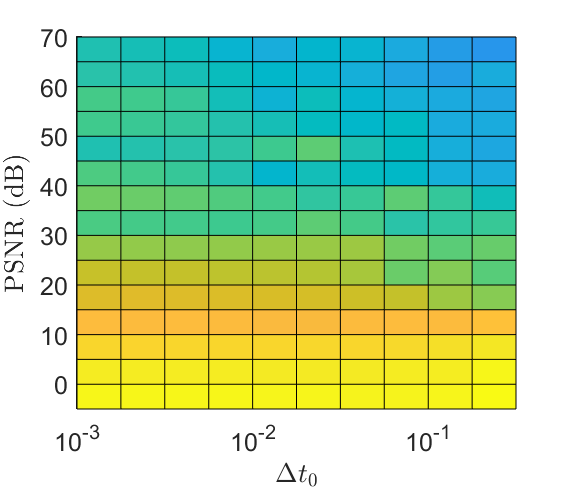}}
\hspace*{-0.4cm}
\includegraphics[width=0.084\linewidth]{figures/colorbar.png}
\caption{Mean standard deviation of the retrieved locations of a stream of Diracs sampled by E-Spline ($P=6, N=21, K=2$) over 10000 realisations at each PSNR-$\Delta t_0$ pair using different methods.
}
\label{fig:ESplineresultsunknown}
\end{figure}
\begin{table*}[!b]
\renewcommand{\arraystretch}{1.6}
\caption{A Comparison between Classical FRI Techniques and Our Approaches in Reconstructing $K=2$ Diracs from $N=21$ Samples}
\label{table:comparisonmethods}
\centering
\begin{tabularx}{\textwidth}{ZYYYY}
\toprule
\noalign{\vspace{-0.5ex}}
& Prony's method with Cadzow denoising \cite{Prony1795, Cadzow1988} & Prony's method with Deep Unfolded PWGD & Direct inference using DNN (Encoder of FRIED-Net) \cite{Leung2020} & FRIED-Net \\ \midrule
Sampling Kernel $\varphi(t)$ & Known ($c_{m,n}$) & Known ($c_{m,n}$) & Not required when training or testing & Fixed decoder when known; Learned via backpropagation when unknown ($d_i$); Not required when testing \\ 
SVD required? & Yes, 1 per iteration & Yes, 1 per layer & No & No \\ 
Number of \emph{free} parameters used & Not applicable & 2425 & 281,002 & \quad 281,002 (Encoder) \newline 1344 (Decoder) \\ %$\left(4(P-M+1)^2+1\right)L$ 
Performance at high PSNR and when Diracs are far apart & Closely follows the Cram\'{e}r-Rao bound & Closely follows the Cram\'{e}r-Rao bound; As good as Cadzow & Provides a rough estimate of the locations hence plateaus; Can be improved coupled with gradient descent when sampling kernel is known & Refines the estimate of the locations from direct inference; Can be improved by fine-tuning the trained model per test datum \\ 
Performance at low PSNR and when Diracs are close together & Breaks down due to so-called subspace swap events & Overcomes breakdown PSNR but slightly erratic & Overcomes breakdown PSNR & Overcomes breakdown PSNR and refines the estimate of the locations from direct inference  \\ 
\bottomrule
\end{tabularx}
\end{table*}
To further show that this network structure is capable of learning any arbitrary kernels other than eMOMS, we repeat the simulation in \cref{sect:eMOMSunknown} using a different sampling kernel. To allow comparison with classical FRI techniques, we choose the sampling kernel to be another exponential reproducing function, that is an E-Spline that can reproduce $P+1=7$ exponentials with $\omega_0 = \frac{-P\pi}{P+1}$ and $\lambda =\frac{2\pi}{3.5(P+1)}$ as the sampling kernel. Same as the previous simulation, we focus on a simple case of having two Diracs with equal amplitudes $a_0 = a_1 \sim \mathcal{U}(0.5,10)$ and evaluate the performance by changing the distance between neighbouring Diracs evenly on a logarithmic scale. 

\cref{fig:ESpline_estimated} shows the estimated sampling kernel obtained from the learned coefficients $\mathbf{d}$. Similar to the case of eMOMS, the network is able to learn the sampling kernel up to PSNR $=10$ dB, while only capturing the main peak together with some oscillations caused by the noise at PSNR $=0$ dB. The reconstruction performance is shown in \cref{fig:ESplineresultsunknown}.

We observe that despite a similar trend, the overall performance is worse than that of eMOMS, regardless of classical or learning-based FRI algorithms. This is as expected because eMOMS is a more effective kernel for FRI recovery than E-Splines \cite{Uriguen2013}. Second, we also see that FRIED-Net alleviates the breakdown inherent to classical subspace-based FRI methods. However, it once again exhibits a compromise in the situation where the noise level is low and the Diracs are sufficiently far apart. This shows that FRIED-Net can reconstruct from discrete samples acquired from kernels other than eMOMS.

\subsection{Summary}

In this section, we summarise and compare our approaches with classical FRI techniques in terms of required information, complexity and reconstruction performance, as well as discussing why learning-based approaches are able to overcome the breakdown events. \cref{table:comparisonmethods} highlights the key findings from the simulation for $K=2$, which also generalises to cases where we reconstruct more pulses. In terms of sampling kernel $\varphi(t)$, as both classical FRI and Deep Unfolded PWGD involves Prony's method, they require this information to be known to translate FRI reconstruction problem into spectral estimation. For FRIED-Net, while the encoder can be trained on its own without the knowledge of $\varphi(t)$, the decoder of FRIED-Net can be either fixed or learned depending on whether it is known. Here $\varphi(t)$ is not required in the evaluation stage as we only need the encoder to reconstruct the locations. Hence, FRIED-Net is more suitable in applications such as calcium imaging when the pulse is unknown, as we will later show in \cref{sect:application}.

In terms of complexity, FRIED-Net involves more than 100 times the number of free parameters used in Deep Unfolded PWGD. However, this is counteracted by the fact that both classical FRI and Deep Unfolded PWGD performs one SVD per iteration, which requires high complexity.

We can then discuss the reconstruction performance by dividing it into two cases: before and after the classical FRI techniques break down. When the PSNR is high and the locations are far apart, both Prony's method with Cadzow and Deep Unfolded PWGD closely follows the Cram\'{e}r-Rao bound, while the performance of FRIED-Net plateaus despite the decoder refining the estimates. For the breakdown region, all of our proposed learning-based algorithms are able to overcome the breakdown PSNR. This is because in general, our learning-based approaches utilises the ground truth labels in the training data effectively as a prior knowledge in the system. In this particular FRI reconstruction problem, classical FRI methods are most vulnerable to pulses that are close together, causing the breakdown events. We conjecture that by feeding the network with training data that contains Diracs with variable distances, including when they are really close together, our proposed learning-based approaches are able to learn from this prior knowledge, and hence the classical performance bound, that is the breakdown PSNR, no longer applies. Amongst our proposed algorithms, the full FRIED-Net provides the best result as it refines the estimation from direct inference and is less erratic than Deep Unfolded PWGD. 

\section{Application to Neuroscience - Calcium Imaging}
\label{sect:application}

In this section, we show how our proposed FRIED-net can be applied to a real life scenario in spike detection from calcium imaging data. Monitoring neural activity has been a key problem to understanding how neural circuits work in animals or humans. As neural activity changes the intracellular calcium concentration \cite{Yasuda2004}, fluorescent calcium sensors offer a way to monitor a large number of cells at the same time. Previous work \cite{Onativia2013} considered that calcium transients model a stream of decaying exponentials and reconstructed the stream using FRI theory. Here, we demonstrate a similar usage of FRIED-Net, yet without explicitly specifying the sampling kernel as a decaying exponential, similar to the simulation in \cref{sect:unknownkernel}.

\subsection{Method}

\subsubsection{Calcium Imaging Dataset}
\begin{figure}[t]
    \centering
    \vspace*{-0.2cm}
    \input{figures/calciumdatasegmentation.tex}
    \caption{Simultaneous imaging (top) and spikes (bottom) of a GCaMP6f expressing neuron from cai-1 dataset \cite{Chen2013, Svoboda2015}. Data segmentation is performed by a sliding window with a step of 1 sample.}
    \label{fig:cai1data}
    \vspace*{-0.3cm}
\end{figure}
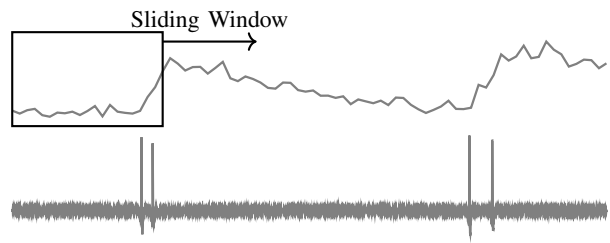

We use the cai-1 dataset \cite{Chen2013, Svoboda2015}, which contains simultaneous imaging with loose-seal cell-attached recording in GCaMP6f expressing neurons. Here, the calcium imaging data is equivalent to the noisy samples $\tilde{y}[n]$, while the simultaneous cell-attached recording provides the ground truth spikes $t_k$ for the training data. Each of the images lasts 240 seconds and is sampled at 60 Hz and the temporal resolution of the spikes is 100 ms. We further choose a subset of 9 recordings from the same cell, where 8 of them would be the training data and the remaining one is the test data. An example is shown in \cref{fig:cai1data}.

\subsubsection{Data Preprocessing and Spike Detection}
Before applying the data to FRIED-Net, we have to pre-process the data to ensure they behave like FRI signals and suitable to be used in DNNs. First, we perform neuropil correction by subtracting a surrounding neuropil signal from the signal in the region of interest. This avoids the data getting contaminated by the surroundings and makes sure that calcium transients are solely caused by the recorded spikes, hence making it like an FRI signal. 

Second, data segmentation is necessary since it is difficult for a neural network to handle such a long and variable-length data stream with a large amount of spikes to be recovered. Hence, we use a sliding window with a moving step of 1 sample to divide the entire data stream into segments. In this way we also effectively increase the amount of training data we have, as each sample is now present in multiple data segments collected by the overlapping windows. After collecting the data segments, we remove the bias in each segment by subtracting the entire window with the smallest sample and rescale the ground truth spikes to $t_k\in[-0.5, 0.5)$ for the usage of FRIED-Net.

\begin{figure*}[t]
    \vspace*{-0.75cm}
    \centering
    \subfloat[Ground truth and recovered spikes of a snippet of the test fluorescent signal using our proposed FRIED-Net (Probability threshold $=0.1$).  \label{fig:calcium_recon}]{
    \centering
    \includegraphics[width=0.55\linewidth]{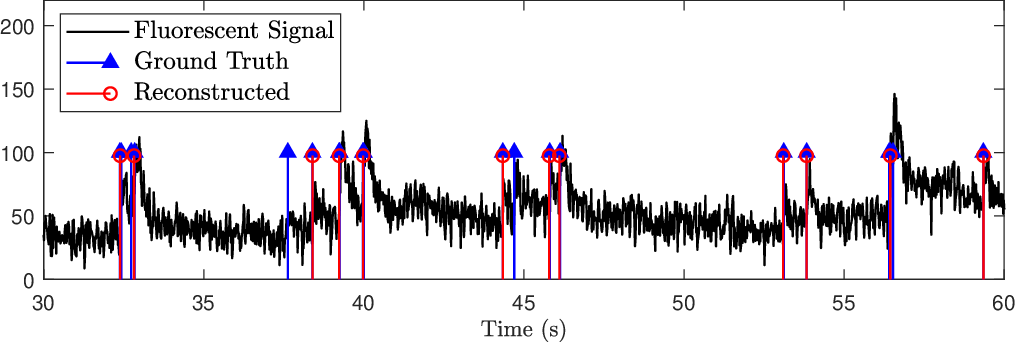}
    }
    \subfloat[ROC curves. \label{fig:roccurves}]{
    \centering
    \includegraphics[width=0.25\linewidth]{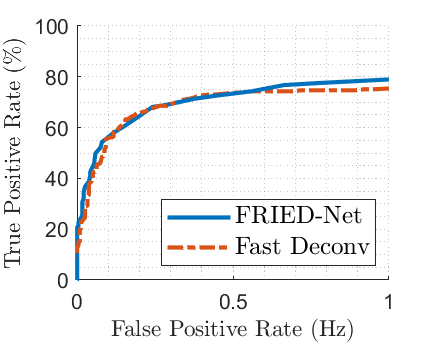}
    }
    \hspace*{-0.45cm}
    \subfloat[Standard deviation of the true positives w.r.t. the ground truth spikes. \label{fig:boxplot}]{
    \hspace*{-0.3cm}
    \includegraphics[width=0.18\linewidth]{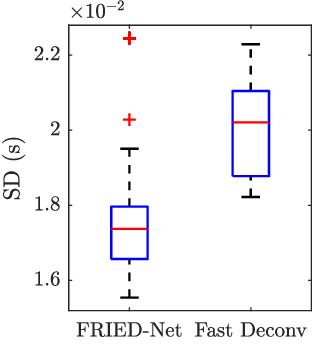}
    }
    \caption{Spike detection performance of our proposed FRIED-Net ($N_{short}=\{32, 16\}, N_{long}=\{128, 64, 32\}, K_{long}=7, T_a = 2T = 0.033$ s) on test data from cai-1 dataset \cite{Chen2013, Svoboda2015}.}
    \vspace*{-0.3cm}
\end{figure*}

Together with data segmentation, we also employ the double consistency approach in \cite{Onativia2013}. Specifically, we run the algorithm with two different strategies: we detect a single spike ($K_{short}=1$) in a sufficiently short window ($N_{short}$) and we detect multiple spikes ($K_{long}$) in a sufficiently long window ($N_{long}$). While the short window is able to provide a precise estimate of a single spike in a small time frame, the long window is able to capture a rough estimate of multiple spikes. When a reconstructed spike corresponds to an actual spike (true positive), its estimated location will be consistent across different windows. Contrarily, as the algorithm treats each window independently, if a spike is found due to noise (false positive), its location estimation will likely be unstable across windows. Therefore, we can collate the outputs from both strategies and construct an aggregated histogram, where the peaks of the histogram give us the candidates of estimated spikes and the magnitude of the peaks (between 0 and 1) provides us the probability of that corresponding to an actual spike. We then threshold the probability to select the probable candidates as our final estimation. The lower this threshold is, the likelier the algorithm achieves a higher detection rate, yet a higher false positive rate at the same time, effectively creating a trade-off between true positive rate and false positive rate. An estimated spike $\hat{t}_k$ is treated as a true positive when it is within the acceptance threshold $T_a$ of an actual spike, i.e. $\hat{t}_k \in [t_k-T_a, t_k+T_a]$, and vice versa.

However, since the ground truth locations have to be labelled in order to learn FRIED-Net, some further tweaks to the data are still necessary. First, in the training data, we only include the windows where spikes exist, since we focus on the accuracy and precision of finding the true positives and the double consistency approach would ideally eliminate any spikes caused by noise in testing. As the data segments likely contain streams of variable number of pulses, the model orders ($K_{short}$ and $K_{long}$) only now specify the \emph{maximum} number of pulses. In the case where $K$ is less than the number of ground truth spikes, we simply choose the first $K$ spikes from the ground truth. In the case of having a sufficiently long window to detect multiple spikes, there exists a possibility that $K$ is larger than the number of actual spikes. Previously in \cite{Onativia2013}, SVD was used to estimate the number of spikes prior to using the FRI algorithms. However, since neural networks are not usually capable of output of variable size, we instead label the ``non-existent'' spikes as arbitrary spikes that are outside of the window. In our case, as the window ranges from -0.5 to 0.5, we set the locations of the arbitrary spikes to be 1. When constructing the histogram, any reconstructed spikes outside the window will be disregarded. Together with the double consistency approach, these tweaks allow us to deal with data containing a variable number of pulses. 

\subsubsection{Training FRIED-Net}
Given the preprocessed data, we simply take the samples from each window as $\{\tilde{y}[n]\}_{n=0}^{N-1}$, feed them into FRIED-Net and estimate the locations of the $K$ spikes $\{\hat{t}_k\}_{k=0}^{K-1}$. We can then learn an individual FRIED-Net for each window length configuration. However, as explained in \cref{sect:10Diracs}, the full FRIED-Net is better at locating a small number of spikes precisely, while the direct inference network (using only the encoder of FRIED-Net) is good at making rough estimates of a high number of spikes. Hence, for the short window, we train the full FRIED-Net and employ the strategy mentioned in \cref{sect:trainingstrat} when the kernel is unknown. The loss function is the combination of squared error on the reconstructed samples and the estimated locations, as stated in \cref{sect:FRIEDlossfunction}. For the long window, we simply train the encoder of FRIED-Net as the direct inference network, with the loss function being the squared error on the locations of the spikes only. 

\subsection{Simulation Results}
\label{sect:calciumresults}

In this section, we present the simulation results of our proposed algorithm on real-life calcium imaging data. As we experimentally found out that better performance is achieved when we run multiple window lengths for each strategy, we set the window lengths for short and long windows to be $N_{short}=\{32, 16\}$ and $N_{long}=\{128, 64, 32\}$ respectively. For the case of long windows, we are recovering $K_{long}=7$ spikes. A network is trained for each window length and the recovered spikes from each network is collated into a histogram as aforementioned. \cref{fig:calcium_recon} shows an example of the reconstructed spikes using FRIED-Net when the probability threshold is $0.1$, overlaid with the ground truth spikes. We observe that our algorithm successfully captures most of the spikes, even when the spikes are close together.

To quantitatively compare our approach against the broadly used probabilistic fast deconvolution algorithm \cite{Vogelstein2010}, we present the receiver operating characteristic (ROC) curves of the respective techniques in \cref{fig:roccurves}. It is plotted by changing the threshold of the probability histogram between 0 and 1 to illustrate the trade-off between the spike detection rate and the false positive rate. Here, the acceptance interval $T_a$ is chosen to be double the sampling period, that is $0.033$ s. We observe that FRIED-Net performs competitively over the fast deconvolution algorithm, as it achieves a true positive rate of $80\%$. We further compare the precision of the true positives. Again, we use the standard deviation described in \eqref{eq:sd} as the evaluation metric. \cref{fig:boxplot} presents the distribution of standard deviation of the detected locations with respect to the ground truth locations with changing probability threshold, where outliers are indicated by the red crosses. We see that the overall standard deviation of FRIED-Net is lower than that of the fast deconvolution algorithm. This shows that our approach provides more precise estimations.

\section{Conclusion}
\label{sect:conclusion}

This paper addresses limitations of existing FRI techniques in that the reconstruction performance breaks down in the presence of noise caused by the so-called subspace swap event. We proposed two learning-based FRI reconstruction algorithms that are inspired by the classical FRI reconstruction models. Deep Unfolded PWGD provides an interpretable deep neural network based on existing iterative denoising algorithm for subspace-based methods, while FRIED-Net aims to bypass subspace-based algorithms and instead models the acquisition process of FRI signals. The latter is particularly useful in neuroscience applications where the sampling kernel $\varphi(t)$ is unknown, since it can be learned using backpropagation. Simulation results show that despite a slight compromise at high PSNR, our proposed approaches reconstruct FRI signals in the low PSNR region where existing FRI algorithms break down, even when the original sampling kernel is unknown. We then demonstrated that our proposed approach provides more precise spike detection than existing algorithms on real-life calcium imaging data, while maintaining a similar performance in terms of true positive and false positive rate. 

% \vspace*{-0.1cm}
\bibliographystyle{IEEEtran}
\bibliography{format,library}

\end{document}

%% file: figures/acquisition.tex
\tikzstyle{block} = [draw, rectangle, minimum height=3em, minimum width=6em]
\tikzstyle{input} = [draw=none]
\tikzstyle{output} = [coordinate]

% The block diagram code is probably more verbose than necessary
\begin{tikzpicture}[auto, node distance=2cm,>=latex']
    % We start by placing the blocks
    \node [input] at (0,0) (input) {$x(t)$};
    \node [block, right of=input, node distance=3cm] (sampler) {$h(t)=\varphi(-t/T)$};
    \node [coordinate, right of=sampler] (output) {};
    \node [coordinate, right of=output, xshift= -1cm](test){};
    \node [draw=none, right of=test, xshift= -1cm](test1){$y[n]$};

    % Once the nodes are placed, connecting them is easy. 
	\draw [->] (input) -- (sampler);
	\draw [-] (sampler) -- (output);
	\draw [-] (output) to[cspst] node[name=t, xshift= -0.5cm, yshift= 0.25cm] {\footnotesize
 $t=nT$} (test);
	\draw [->] (test) -- (test1);

\end{tikzpicture}

%% file: figures/unfoldedwirtinger.tex
\tikzstyle{block} = [draw, rounded corners, rectangle, minimum height=3em, minimum width=3em]

\def\textcolorblue{rgb:red,200;green,20;blue,20}
\def\sep{2.25cm}
\def\vsep{3cm}
% The block diagram code is probably more verbose than necessary
\begin{tikzpicture}[auto, node distance=2*\sep/3, >=latex']
    % We start by placing the blocks
    \node [block] (L0) {$\mathbf{L}^{(0)}$};
    \node [block, below of=L0, node distance =\vsep] (H0) {$\mathbf{H}^{(0)}$};
    \node [circle, draw, right of=L0, node distance =0.75*\sep] (1+) {$+$};
    \node [circle, draw, right of=H0, node distance =1.1*\sep] (1++) {$+$};
    \node [circle, draw, right of=1++, minimum size = 1.75cm] (1PT) {$\mathcal{P}_{\mathcal{T}}$};
    
    \node [circle, draw, above of=1PT, node distance =\vsep, minimum size = 1.75cm] (1P) {$\mathcal{S}_{\textcolor[RGB]{200,20,20}{\mu^{(0)}}\sigma_{K+1}}$};
    \node [block, right of=1P, node distance =\sep-0.5cm] (L1) {$\mathbf{L}^{(1)}$};
    
    \node [block, right of=1PT, node distance =\sep-0.5cm] (H1) {$\mathbf{H}^{(1)}$};
    \node [right of=L1, node distance =\sep/2] (Ldots) {$\cdots$};
    \node [right of=H1, node distance =\sep/2] (Hdots) {$\cdots$};
    \node [block, right of=Ldots, node distance =\sep/2] (LL) {$\mathbf{L}^{(L)}$};
    \node [block, right of=Hdots, node distance =\sep/2] (HL) {$\mathbf{H}^{(L)}$};

    % Once the nodes are placed, connecting them is easy. 
    \draw[->](L0) -- node[text=\textcolorblue]{$\mathbf{W}_1^{(0)}$}(1+);
    \draw[->](1+) -- node {}(1P);
    \draw[->](1P) -- node {}(L1);
    % \draw[->](1+)++(0,-\vsep) -- node[text=\textcolorblue]{$\mathbf{W}_2^{(0)}$} (1+);
    % \draw[->](H0) -- node[text=\textcolorblue, near end]{$\mathbf{W}_4^{(0)}$} (1++);
    \draw[->](H0) -- ++(0,\vsep/2) -| node[text=\textcolorblue, above, near start]{$\mathbf{W}_2^{(0)}$} (1+);
    \draw[->](H0) -- node[text=\textcolorblue]{$\mathbf{W}_4^{(0)}$} (1++);
    \draw[->](L1) -- ++(0,-\vsep/2) -| node[text=\textcolorblue, above, near start]{$\mathbf{W}_3^{(0)}$} (1++);
    \draw[->](1++) -- node {}(1PT);
    \draw[->](1PT) -- node {}(H1);
    \draw[->](L1) -- node {}(Ldots.west);
    \draw[->](H1) -- node {}(Hdots.west);
    \draw[->](Ldots.east) -- node {}(LL);
    \draw[->](Hdots.east) -- node {}(HL);
\end{tikzpicture}

%% file: figures/autoencoder_structure.tex
\tikzset{trapezium stretches=true}

\def\blockheight{15mm}
\def\blockwidth{30mm}
\def\layersep{5mm}
\def\codesep{0.45cm}
\def\samplesep{0.75cm}
\begin{tikzpicture}

\foreach [count=\k from 0] \name in {$\hat{t}_0$,$\hat{t}_1$,$\vphantom{\int\limits^x}\smash{\vdots}$,$\hat{t}_{K-1}$}
    \path[yshift=0cm]
        node (t-\k) at (0,-\codesep*\k) {\name};

\node [draw, trapezium, minimum width=\blockwidth, minimum height=\blockheight,shape border rotate=270,trapezium right angle=82,trapezium left angle=82,left=\layersep of t-0,anchor=east,inner sep=0pt,yshift=-\codesep*3/2, fill=red!10] (E) {\begin{tabular}{c} Encoder \\ $g_\phi$ \end{tabular}};

\node [draw, trapezium, minimum width=\blockwidth, minimum height=\blockheight,shape border rotate=90,trapezium right angle=82,trapezium left angle=82,right=\layersep of t-0,anchor=west,inner sep=0pt,yshift=-\codesep*3/2, fill=blue!10] (D) {\begin{tabular}{c} Decoder \\ $f_\theta$ \end{tabular}};
\foreach [count=\n from 0] \name in {$\tilde{y}[0]$,$\tilde{y}[1]$,$\vphantom{\int\limits^x}\smash{\vdots}$,$\tilde{y}[N-1]$}
    \path[yshift=(\samplesep-\codesep)*3/2]
        node[left= 5.5*\layersep of E] (y-\n) at (0,-\samplesep*\n) {\name};

\foreach [count=\n from 0] \name in {$\hat{y}[0]$,$\hat{y}[1]$,$\vphantom{\int\limits^x}\smash{\vdots}$,$\hat{y}[N-1]$}
    \path[yshift=(\samplesep-\codesep)*3/2]
        node[right= 5.5*\layersep of D] (yhat-\n) at (0,-\samplesep*\n) {\name};

\foreach \k in {0,1,3}{
    \draw[<-] (t-\k)++(-\layersep,0) -- (t-\k-|E.east);;
    \draw[->] (t-\k)++(\layersep,0) -- (t-\k-|D.west);
}

\foreach \n in {0,1,3}{
    \draw[->] (y-\n) -- (y-\n-|E.west);
    \draw[<-] (yhat-\n) -- (yhat-\n-|D.east);
}
\node[above=0.6cm of D] (a-k) {$\{\hat{a}_k\}_{k=0}^{K-1}$};
\draw[->] (a-k) -- (D);

\end{tikzpicture}

%% file: figures/decoder_architecture.tex
\def\layersep{1.5cm}

\begin{tikzpicture}[shorten >=1pt,->,draw=black!50, node distance=\layersep]
    \tikzstyle{every pin edge}=[<-,shorten <=1pt]
    \tikzstyle{neuron}=[draw=black,circle,minimum size=17pt,inner sep=0pt]
    \tikzstyle{input neuron}=[neuron, fill=black!10];
    \tikzstyle{output neuron}=[neuron, fill=black!10];
    \tikzstyle{hidden neuron}=[neuron, fill=black!10];
    \tikzstyle{relu neuron}=[draw=black,minimum size=17pt,inner sep=2pt, fill=black!10, path picture={% 
        \draw[thick, -] (0.5em,0.5em) -- (0,-0.25em) -- (-0.5em,-0.25em);
      }];
    \tikzstyle{annot} = [text width=4em, text centered]

    % Draw the input layer nodes
    \foreach [count=\x from 0] \name / \y in {1}
    % This is the same as writing \foreach \name / \y in {1/1,2/2,3/3,4/4}
        \node[input neuron, pin=left:$\hat{t}_\x$] (I-\name) at (-1,-3 cm) {};

    % Draw the hidden layer nodes
    \foreach \name / \y in {1,...,2}
        \path[yshift=1.5cm]
            node[hidden neuron] (FC1-\name) at (0*\layersep,-3*\y cm) {};

    \foreach \name / \y in {1,...,6}{
        \path[yshift=0.5cm]
            node[relu neuron] (ReLU-\name) at (1*\layersep,-\y cm) {};
        }

    % Draw the output layer node
    \foreach \name / \y in {1,...,2}
        \path[yshift=1.5cm]
            node[output neuron] (Output-\name) at (2*\layersep,-3*\y cm) {};

    \foreach [count=\x from 0] \name / \y in {1,...,2}
        \path[yshift=1.5cm]
            node (Phi-\name) at (3.05*\layersep,-3*\y cm) {$\hat{\varphi}_\Delta\left(\frac{\hat{t}_0}{T}-\x\right)$};

    \foreach [count=\x from 0] \name / \y in {1,...,2}
        \path[yshift=1.5cm]
            node (y-\name) at (4.15*\layersep,-3*\y cm) {$\hat{y}[\x]$};

    % Connect every node in the input layer with every node in the
    % hidden layer.
    \foreach \source in {1}
        \foreach [count=\dest from 1]\x in {0,...,-1}
            \draw[->] (I-\source) -- node[above,sloped,font=\bfseries]{\footnotesize 1/T} node[below,sloped]{\footnotesize\x} ++(FC1-\dest);
            
    \foreach \source in {1,...,2}
        \foreach [count=\dest from \source*3-2, count = \i from 1] \x in {0,-1,-2}
            \draw[->] (FC1-\source) -- node[below]{\footnotesize\x$\Delta$} ++(ReLU-\dest);

    % \foreach \source in {1,...,9}
    %     \path (FC2-\source) edge (ReLU-\source);

    \foreach \dest in {1,...,2}
        \foreach [count=\source from \dest*3-2, count=\x from 0] \i in {$\bf d_0$,$\bf d_1$,$\bf d_2$}
            \draw[->] (ReLU-\source) -- node[above,font=\bfseries]{\footnotesize\i} ++(Output-\dest);

    \foreach \source in {1,...,2}
        \path (Output-\source) edge (Phi-\source);
        
    \foreach \source in {1,...,2}
        \path (Phi-\source) edge ["{\footnotesize $\bf \hat{a}_0$}"] (y-\source);

    % Annotate the layers
    \node[annot,above of=ReLU-1, node distance=0.75cm] (fc2) {FC2+ReLU};
    \node[annot,left of=fc2] (fc1) {FC1};
    \node[annot,left of=fc1] {Input};
    \node[annot,right of=fc2, node distance=2*\layersep] {Output};
    
    % Legend
   \matrix [draw,matrix of nodes,nodes={anchor=west},below right] at ([yshift=-1mm]current bounding box.south west) {
      $x$ & \draw [->] (-2,0) -- node[above,sloped,font=\bfseries] {\footnotesize Weight $\bf w$} node[midway,below] {\footnotesize Bias $b$} ++(2,0); & \node[input neuron, pin={[pin edge={->}]right:$\mathbf{w}\cdot x+b$}] at (1.5,0) {}; \\
    };
\end{tikzpicture}

%% file: figures/calciumdatasegmentation.tex
\begin{tikzpicture}[auto]

    \draw[gray, line width=0.3mm] plot file {figures/calciumdata.txt};
    \draw[gray, line width=0.3mm] plot file {figures/ephysdata.txt};

    % \draw[|-|, thick] (0,-0.5) -- node[below]{\small 1st Window} (2,-0.5);    
    % \draw[|-|, thick] (0.1,-1) -- node[below]{\small 2nd Window} (2.1,-1);   
    
    % Square sliding window, arrow to the right
    \draw[draw=black, thick] (0,-0.125) rectangle ++(2,1.25);
    \draw[->, thick] (2,1) -- node[above]{\small Sliding Window} (3.25,1);
    
\end{tikzpicture}

%% file: main.bbl
\begin{thebibliography}{10}
\providecommand{\url}[1]{#1}
\csname url@samestyle\endcsname
\providecommand{\newblock}{\relax}
\providecommand{\bibinfo}[2]{#2}
\providecommand{\BIBentrySTDinterwordspacing}{\spaceskip=0pt\relax}
\providecommand{\BIBentryALTinterwordstretchfactor}{4}
\providecommand{\BIBentryALTinterwordspacing}{\spaceskip=\fontdimen2\font plus
\BIBentryALTinterwordstretchfactor\fontdimen3\font minus
  \fontdimen4\font\relax}
\providecommand{\BIBforeignlanguage}[2]{{%
\expandafter\ifx\csname l@#1\endcsname\relax
\typeout{** WARNING: IEEEtran.bst: No hyphenation pattern has been}%
\typeout{** loaded for the language `#1'. Using the pattern for}%
\typeout{** the default language instead.}%
\else
\language=\csname l@#1\endcsname
\fi
#2}}
\providecommand{\BIBdecl}{\relax}
\BIBdecl

\bibitem{Eldar2014}
Y.~C. Eldar, \emph{Sampling Theory: Beyond Bandlimited Systems}.\hskip 1em plus
  0.5em minus 0.4em\relax {Cambridge University Press}, 2014.

\bibitem{Vetterli2002}
M.~Vetterli, P.~Marziliano, and T.~Blu, ``Sampling signals with finite rate of
  innovation,'' \emph{IEEE Transactions on Signal Processing}, vol.~50, no.~6,
  pp. 1417--1428, Jun. 2002.

\bibitem{Dragotti2007}
P.~L. Dragotti, M.~Vetterli, and T.~Blu, ``Sampling moments and reconstructing
  signals of finite rate of innovation: {{Shannon}} meets {{Strang-Fix}},''
  \emph{IEEE Transactions on Signal Processing}, vol.~55, no. 5 I, pp.
  1741--1757, 2007.

\bibitem{Blu2008}
T.~Blu, P.~L. Dragotti, M.~Vetterli, P.~Marziliano, and L.~Coulot, ``Sparse
  sampling of signal innovations: Theory, algorithms, and performance bounds,''
  \emph{IEEE Signal Processing Magazine}, vol.~25, no.~2, pp. 31--40, 2008.

\bibitem{Uriguen2013}
J.~A. Urig{\"u}en, T.~Blu, and P.~L. Dragotti, ``{{FRI}} sampling with
  arbitrary kernels,'' \emph{IEEE Transactions on Signal Processing}, vol.~61,
  no.~21, pp. 5310--5323, 2013.

\bibitem{Tur2011}
R.~Tur, Y.~C. Eldar, and Z.~Friedman, ``Innovation rate sampling of pulse
  streams with application to ultrasound imaging,'' \emph{IEEE Transactions on
  Signal Processing}, vol.~59, no.~4, pp. 1827--1842, 2011.

\bibitem{Onativia2013}
J.~O{\~n}ativia, S.~R. Schultz, and P.~L. Dragotti, ``A finite rate of
  innovation algorithm for fast and accurate spike detection from two-photon
  calcium imaging,'' \emph{Journal of Neural Engineering}, vol.~10, no.~4, pp.
  46\,017--46\,031, 2013.

\bibitem{Dogan2014}
Z.~Do{\u g}an, T.~Blu, and D.~Van De~Ville, ``Detecting spontaneous brain
  activity in functional magnetic resonance imaging using finite rate of
  innovation,'' in \emph{2014 {{IEEE International Symposium}} on {{Biomedical
  Imaging}} ({{ISBI}})}, Jul. 2014, pp. 1047--1050.

\bibitem{Bar-Ilan2014}
O.~{Bar-Ilan} and Y.~C. Eldar, ``Sub-{{Nyquist}} radar via doppler focusing,''
  \emph{IEEE Transactions on Signal Processing}, vol.~62, no.~7, pp.
  1796--1811, 2014.

\bibitem{Wagner2012}
N.~Wagner, Y.~C. Eldar, and Z.~Friedman, ``Compressed beamforming in ultrasound
  imaging,'' \emph{IEEE Transactions on Signal Processing}, vol.~60, no.~9, pp.
  4643--4657, 2012.

\bibitem{Hao2005}
Y.~Hao, P.~Marziliano, M.~Vetterli, and T.~Blu, ``Compression of {{ECG}} as a
  signal with finite rate of innovation,'' in \emph{Annual {{International
  Conference}} of the {{IEEE Engineering}} in {{Medicine}} and {{Biology}}},
  vol.~7, 2005, pp. 7564--7567.

\bibitem{Prony1795}
R.~Prony, ``Essai exp\'erimental et analytique sur les lois de la
  dilatabilit\'e des fluides \'elastiques, et sur celles de la force expansive
  de la vapeur de l'eau et de la vapeur de l'alkool, \'a diff\'erentes
  temperatures,'' \emph{J. de l'Ecole Polytechnique}, vol.~1, pp. 24--76, 1795.

\bibitem{Cadzow1988}
J.~A. Cadzow, ``Signal enhancement - {{A}} composite property mapping
  algorithm,'' \emph{IEEE Transactions on Acoustics, Speech, and Signal
  Processing}, vol.~36, no.~1, pp. 49--62, 1988.

\bibitem{Hua1990}
Y.~Hua and T.~K. Sarkar, ``Matrix pencil method for estimating parameters of
  exponentially damped/undamped sinusoids in noise,'' \emph{IEEE Transactions
  on Acoustics, Speech, and Signal Processing}, vol.~38, no.~5, pp. 814--824,
  May 1990.

\bibitem{Cramer1946}
H.~Cram{\'e}r, \emph{Mathematical Methods of Statistics}.\hskip 1em plus 0.5em
  minus 0.4em\relax {Princeton university press}, 1946.

\bibitem{Rao1945}
C.~R. Rao, ``Information and the accuracy attainable in the estimation of
  statistical parameters,'' \emph{Bulletin of Calcutta Mathematical Society},
  vol.~37, pp. 81--89, 1945.

\bibitem{Wei2015}
X.~Wei and P.~L. Dragotti, ``Guaranteed performance in the {{FRI}} setting,''
  \emph{IEEE Signal Processing Letters}, vol.~22, no.~10, pp. 1661--1665, 2015.

\bibitem{Thomas1995}
J.~K. Thomas, L.~L. Scharf, and D.~W. Tufts, ``The probability of a subspace
  swap in the {{SVD}},'' \emph{IEEE Transactions on Signal Processing},
  vol.~43, no.~3, pp. 730--736, 1995.

\bibitem{Bhaskar2013}
B.~N. Bhaskar, G.~Tang, and B.~Recht, ``Atomic norm denoising with applications
  to line spectral estimation,'' \emph{IEEE Transactions on Signal Processing},
  vol.~61, no.~23, pp. 5987--5999, 2013.

\bibitem{Tang2013}
G.~Tang, B.~N. Bhaskar, P.~Shah, and B.~Recht, ``Compressed sensing off the
  grid,'' \emph{IEEE Transactions on Information Theory}, vol.~59, no.~11, pp.
  7465--7490, 2013.

\bibitem{Candes2014}
E.~J. Cand{\`e}s and C.~{Fernandez-Granda}, ``Towards a mathematical theory of
  super-resolution,'' \emph{Communications on Pure and Applied Mathematics},
  vol.~67, no.~6, pp. 906--956, 2014.

\bibitem{Mulleti2020}
S.~Mulleti, K.~Lee, and Y.~C. Eldar, ``Identifiability conditions for
  compressive multichannel blind deconvolution,'' \emph{IEEE Transactions on
  Signal Processing}, vol.~68, pp. 4627--4642, 2020.

\bibitem{Tolooshams2022}
B.~Tolooshams, S.~Mulleti, D.~Ba, and Y.~C. Eldar, ``Learning filter-based
  compressed blind-deconvolution,'' \emph{arXiv:2209.14165}, Sep. 2022.

\bibitem{Mathew1994}
G.~Mathew and V.~U. Reddy, ``Development and analysis of a neural network
  approach to pisarenko's harmonic retrieval method,'' \emph{IEEE Transactions
  on Signal Processing}, vol.~42, no.~3, pp. 663--667, 1994.

\bibitem{Izacard2019}
G.~Izacard, S.~Mohan, and C.~{Fernandez-Granda}, ``Data-driven estimation of
  sinusoid frequencies,'' in \emph{Advances in Neural Information Processing
  Systems}, vol.~32, 2019, pp. 5127--5137.

\bibitem{Izacard2019a}
G.~Izacard, B.~Bernstein, and C.~{Fernandez-Granda}, ``A learning-based
  framework for line-spectra super-resolution,'' in \emph{2019 {{IEEE
  International Conference}} on {{Acoustics}}, {{Speech}} and {{Signal
  Processing}} ({{ICASSP}})}.\hskip 1em plus 0.5em minus 0.4em\relax {IEEE},
  May 2019, pp. 3632--3636.

\bibitem{Adavanne2018}
S.~Adavanne, A.~Politis, and T.~Virtanen, ``Direction of arrival estimation for
  multiple sound sources using convolutional recurrent neural network,'' in
  \emph{2018 {{European Signal Processing Conference}} ({{EUSIPCO}})}, Nov.
  2018, pp. 1462--1466.

\bibitem{Xiao2015}
X.~Xiao \emph{et~al.}, ``A learning-based approach to direction of arrival
  estimation in noisy and reverberant environments,'' in \emph{2015 {{IEEE
  International Conference}} on {{Acoustics}}, {{Speech}} and {{Signal
  Processing}} ({{ICASSP}})}, 2015, pp. 2814--2818.

\bibitem{Shlezinger2021}
N.~Shlezinger, J.~Whang, Y.~C. Eldar, and A.~G. Dimakis, ``Model-based deep
  learning: Key approaches and design guidelines,'' in \emph{2021 {{IEEE Data
  Science}} and {{Learning Workshop}}, {{DSLW}} 2021}, Jun. 2021.

\bibitem{Gregor2010}
K.~Gregor and Y.~LeCun, ``Learning fast approximations of sparse coding,'' in
  \emph{2010 {{International Conference}} on {{Machine Learning}} ({{ICML}})},
  2010, pp. 399--406.

\bibitem{Monga2021}
V.~Monga, Y.~Li, and Y.~C. Eldar, ``Algorithm unrolling: Interpretable,
  efficient deep learning for signal and image processing,'' \emph{IEEE Signal
  Processing Magazine}, vol.~38, no. March, pp. 18--44, 2021.

\bibitem{Solomon2020}
O.~Solomon \emph{et~al.}, ``Deep unfolded robust {{PCA}} with application to
  clutter suppression in ultrasound,'' \emph{IEEE Transactions on Medical
  Imaging}, vol.~39, no.~4, pp. 1051--1063, 2020.

\bibitem{Cai2015a}
J.-F. Cai, S.~Liu, and W.~Xu, ``A fast algorithm for reconstruction of
  spectrally sparse signals in super-resolution,'' in \emph{Wavelets and
  {{Sparsity XVI}}}, vol. 9597.\hskip 1em plus 0.5em minus 0.4em\relax {SPIE},
  Aug. 2015, p. 95970A.

\bibitem{Dang2018}
Z.~Dang \emph{et~al.}, ``Eigendecomposition-free training of deep networks with
  zero eigenvalue-based losses,'' in \emph{2018 {{European Conference}} on
  {{Computer Vision}} ({{ECCV}})}, 2018, pp. 768--783.

\bibitem{Leung2021}
V.~C.~H. Leung, J.-J. Huang, Y.~C. Eldar, and P.~L. Dragotti, ``Reconstruction
  of {{FRI}} signals using autoencoders with fixed decoders,'' in \emph{2021
  {{European Signal Processing Conference}} ({{EUSIPCO}})}, 2021, pp.
  1496--1500.

\bibitem{Unser2005}
M.~Unser and T.~Blu, ``Cardinal exponential splines: Part {{I}} - theory and
  filtering algorithms,'' \emph{IEEE Transactions on Signal Processing},
  vol.~53, no.~4, pp. 1425--1438, 2005.

\bibitem{Markovsky2008}
I.~Markovsky, ``Structured low-rank approximation and its applications,''
  \emph{Automatica}, vol.~44, no.~4, pp. 891--909, Apr. 2008.

\bibitem{Fazel2001}
M.~Fazel, H.~Hindi, and S.~P. Boyd, ``A rank minimization heuristic with
  application to minimum order system approximation,'' in \emph{2001 {{American
  Control Conference}} ({{ACC}})}, vol.~6, 2001, pp. 4734--4739.

\bibitem{Condat2015}
L.~Condat and A.~Hirabayashi, ``Cadzow denoising upgraded: A new projection
  method for the recovery of dirac pulses from noisy linear measurements,''
  \emph{Sampling Theory in Signal and Image Processing}, vol.~14, no.~1, pp.
  17--47, 2015.

\bibitem{Kingma2015}
D.~P. Kingma and J.~Ba, ``Adam: {{A Method}} for {{Stochastic Optimization}},''
  in \emph{2015 {{International Conference}} on {{Learning Representations}}
  ({{ICLR}})}, {San Diego, CA, USA}, 2015.

\bibitem{Leung2020}
V.~C.~H. Leung, J.-J. Huang, and P.~L. Dragotti, ``Reconstruction of {{FRI}}
  signals using deep neural network approaches,'' in \emph{2020 {{IEEE
  International Conference}} on {{Acoustics}}, {{Speech}} and {{Signal
  Processing}} ({{ICASSP}})}, 2020, pp. 5430--5434.

\bibitem{Hornik1989}
K.~Hornik, M.~Stinchcombe, and H.~White, ``Multilayer feedforward networks are
  universal approximators,'' \emph{Neural Networks}, vol.~2, no.~5, pp.
  359--366, 1989.

\bibitem{Leshno1993}
M.~Leshno, V.~Y. Lin, A.~Pinkus, and S.~Schocken, ``Multilayer feedforward
  networks with a nonpolynomial activation function can approximate any
  function,'' \emph{Neural Networks}, vol.~6, no.~6, pp. 861--867, 1993.

\bibitem{Shaham2018}
U.~Shaham, A.~Cloninger, and R.~R. Coifman, ``Provable approximation properties
  for deep neural networks,'' \emph{Applied and Computational Harmonic
  Analysis}, vol.~44, no.~3, pp. 759--773, Sep. 2018.

\bibitem{Paszke2019}
A.~Paszke \emph{et~al.}, ``{{PyTorch}}: An imperative style, high-performance
  deep learning library,'' in \emph{Advances in {{Neural Information Processing
  Systems}}}, vol.~32.\hskip 1em plus 0.5em minus 0.4em\relax {Curran
  Associates, Inc.}, 2019, pp. 8024--8035.

\bibitem{Yasuda2004}
R.~Yasuda \emph{et~al.}, ``Imaging calcium concentration dynamics in small
  neuronal compartments.'' \emph{Science's STKE : signal transduction knowledge
  environment}, vol. 2004, no. 219, 2004.

\bibitem{Chen2013}
T.~W. Chen \emph{et~al.}, ``Ultrasensitive fluorescent proteins for imaging
  neuronal activity,'' \emph{Nature}, vol. 499, no. 7458, pp. 295--300, 2013.

\bibitem{Svoboda2015}
K.~Svoboda, ``Simultaneous imaging and loose-seal cell-attached electrical
  recordings from neurons expressing a variety of genetically encoded calcium
  indicators,'' GENIE Project, Janelia Farm Campus, 2015, {CRCNS.org}.

\bibitem{Vogelstein2010}
J.~T. Vogelstein \emph{et~al.}, ``Fast nonnegative deconvolution for spike
  train inference from population calcium imaging,'' \emph{Journal of
  Neurophysiology}, vol. 104, no.~6, pp. 3691--3704, Dec. 2010.

\end{thebibliography}
